\documentclass[utf8]{frontiersSCNS} 

\usepackage{hyperref,lineno,microtype,subcaption}
\usepackage[onehalfspacing]{setspace}
\usepackage{soul}

\def\keyFont{\fontsize{8}{11}\helveticabold }
\def\firstAuthorLast{Panda {et~al.}} 
\def\Authors{Swayamtrupta Panda\,$^{1,2}$, Mary Loli Mart\'inez-Aldama\,$^{1,*}$ and Michal Zaja\v{c}ek\,$^{1}$\\\textit{on behalf of the LSST AGN Science Collaboration}}
\def\Address{$^{1}$Center for Theoretical Physics, Polish Academy of Sciences, Al. Lotnik\'ow 32/46, 02-668 Warsaw, Poland \\
$^{2}$Copernicus Astronomical Center, Polish Academy of Sciences,  ul. Bartycka 18, 00-716 Warsaw, Poland  }
\def\corrAuthor{Swayamtrupta Panda}

\def\corrEmail{panda@cft.edu.pl}

\newenvironment{myfont}{\fontfamily{lmtt}\selectfont}{\par}
\DeclareTextFontCommand{\textmyfont}{\myfont}

\def\kms{\,km\,s$^{-1}$}
\def\hb{H$\beta$}
\def\RL{$R\mathrm{_{H\beta}}-L_{5100}$}
\def\MgII{Mg {\sc ii}}
\def\DL{$D{\mathrm{_L}}$}

\def\fblrc{$f\mathrm{_{BLR}^{\,c}}$}
\def\mdotc{$\dot{{M}}\mathrm{^{c}}$}
\def\DRhb{$\Delta R_{\mathrm{H\beta}}$}

\def\LD{$D\mathrm{_{L}}$}

\newcommand{\angstrom}{\text{\normalfont\AA}}

\begin{document}
\onecolumn
\firstpage{1}

\title[Reverberation-mapped quasars in Cosmology]{Current and future applications of Reverberation-mapped quasars in Cosmology} 

\author[\firstAuthorLast ]{\Authors} 
\address{} 
\correspondance{} 

\extraAuth{}

\maketitle

\begin{abstract}
Reverberation mapping technique (RM) is an important milestone that has elevated our understanding of Active Galactic Nuclei (AGN) demographics, giving information about the kinematics and the structure of the Broad Line Region (BLR). It is based on the time-delay response between the continuum and the emission line. The time delay is directly related to the size of the BLR which in turn is related to the continuum luminosity of the source, producing the well-known Radius-Luminosity (RL) relation.  The majority of the sources with RM data, have been monitored for their \hb\ emission line in low redshift sources ($z<0.1$), while there are some attempts using the \MgII\ line for higher redshift ranges. In this work, we present a recent \MgII\ monitoring for the quasar CTS C30.10 ($z=0.90$) observed with the 10-meter \textit{Southern African Large Telescope (SALT)}, for which the RL scaling based on \MgII\ holds within measurement and time-delay uncertainties. One of the most important advantages of reverberation mapping technique is the independent determination to the distant source, and considering the large range of redshifts and luminosities found in AGNs their use in cosmological studies is promising. However, recently it has been found that highly accreting sources show the time delays shorter than expected from the RL relation. We have proposed a correction for this effect using a sample of 117 \hb\ reverberating-mapped AGN with $0.02<z<0.9$, which recovers the low scatter along with the relation. We are able to determine the cosmological constants, $\Omega_m$ and $\Omega_\Lambda$. Despite the applied correction, the scatter is still large for being effective for cosmological applications. In the near future, \textit{Large Synoptic Survey Telescope (LSST)}\footnote{for more details see \href{https://www.lsst.org/}{https://www.lsst.org/}} will cover over 10 million quasars in six photometric bands during its 10-year run. We present the first step in modelling of light curves for \hb\ and \MgII\ and discuss the quasar selection in the context of photometric reverberation mapping with \textit{LSST}. With the onset of the \textit{LSST} era, we expect a huge rise in the overall quasar counts and redshift range covered ($z\lesssim7.0$), which will provide a better constraint of AGN properties with cosmological purposes. 

\tiny
 \keyFont{ \section{Keywords:} quasars: reverberation mapping, time-delay measurements; cosmology; instrument: LSST; photometry; spectroscopy} 
\end{abstract}

\section{Introduction}

Reverberation mapping (hereafter RM) is a robust observational technique to determine the time-lag $\tau_0$ between the variability of the ionizing continuum of an active galactic nucleus (AGN) and the line emission associated with the broad-line region \citep[BLR, ][]{1982ApJ...255..419B}. This is only possible since the emission-line flux densities are highly correlated with the continuum flux, which implies that the AGN continuum related to the power generated by the accretion disk is the main source of photoionization. The main source of recombination emission lines is the BLR material, that is optically thick with respect to the ionizing UV/optical continuum. 

The first straightforward application of RM is to determine the size of the BLR, $R_{\rm BLR}=c\tau_0$ \citep[see e.g.][]{1997ASSL..218...85N,2000ApJ...533..631K,2004ApJ...613..682P,mejia-restrepo2018}.

The second application combines $\tau_0$ and the line width of the broad emission line. It can be considered that the large line width of the BLR gas $\Delta v_{\rm FWHM}$\footnote{where FWHM is the full width at half-maximum of the corresponding emission line profile} of several thousand km/s, is due to the Doppler broadening of clouds that are gravitationally bound to the supermassive black hole (SMBH). The 3D Keplerian velocity of the gas $v_{\rm BLR}=f_{\rm vir}  \Delta v_{\rm FWHM}$ can be inferred, though it depends on the rather uncertain virial factor $f_{\rm vir}$. By combining these two pieces of information, one can estimate the virial black hole mass,

\begin{equation}
    M_{\bullet}=\frac{R_{\rm BLR}v_{\rm BLR}^2}{G}=f_{\rm vir}\frac{c\tau_0 v_{\rm FWHM}^2}{G}\,,
    \label{eq_bh_mass}
\end{equation}
where the virial factor $f_{\rm vir}$ is of the order of unity and depends on the overall geometrical distribution of the BLR clouds, their kinematics as well as their line of sight emission properties, \textit{c} is the velocity of light and \textit{G} is the Gravitational constant. The virial factor mainly depends on the inclination $i$ of the BLR plane with respect to the observer and the thickness of the BLR, $H_{\rm BLR}/R_{\rm BLR}$, and can be expressed as \citep{collin2006, mejia-restrepo2018, panda19b},

\begin{equation}
    f_{\rm vir}=[4(\sin^2{i}+(H_{\rm BLR}/R_{\rm BLR})^2)]^{-1}\,.
    \label{eq_fvir_geometry}
\end{equation}

Therefore, the virial factor dependence on the viewing angle of the BLR and its geometrical properties, introduces an overall uncertainty in the virial mass determination. Fixing the virial factor to a constant value, which is frequently applied to single-epoch measurements \citep{2015ApJ...801...38W}, can lead to a virial-mass difference of a factor of 2-3 \citep{mejia-restrepo2018}. Comparing the black hole mass estimations determined from the spectral energy distribution (SED) fitting of the AGN continuum, with the viral black hole mass, \citet{mejia-restrepo2018} found that the virial factor is inversely proportional to the FWHM of the broad lines, $f_{\rm vir}\propto \text{FWHM}_{\rm line}^{-1}$. This could be interpreted as the effect of the BLR viewing angle (inclination) or as the effect of the radiation pressure on the BLR distribution. The strong effect of the BLR inclination directly results from the plane-like geometry of the BLR, in which the BLR consists of cloudlets that have an overall flattened geometry \citep{2009NewAR..53..140G}. This ``nest-like'' model of the BLR is also consistent with the first direct velocity-resolved observation of the BLR in 3C273 \citep{2018Natur.563..657G} using the Very Large Telescope Interferometer GRAVITY at the European Southern Observatory \citep{2017A&A...602A..94G}. The flattened geometry of the BLR that follows the disc geometry could be explained by the formation of the BLR clouds from the disc material beyond the dust sublimation radius \citep{2011A&A...525L...8C}. The radiation pressure acting on the dust in the BLR clouds then leads to their lift-off from the disc plane and subsequent fall-back when the dust evaporates. This is one possible scenario for the origin of the low-ionization line component of the BLR (LIL), so-called failed radiatively accelerated disc outflow \citep[FRADO, ][]{2017ApJ...846..154C}, while the high-ionization line (HIL) component seems to be associated with nuclear outflows \citep{1988MNRAS.232..539C}. 

Finally, the third application of the RM is the power-law radius--luminosity relation $R_{\rm BLR}\propto L_{\rm AGN}^{\alpha}$ between the BLR radius (or time-delay) and the AGN monochromatic optical luminosity. Initially, \citet{2005ApJ...629...61K} found the slope $\alpha=0.67\pm 0.05$ between the optical monochromatic luminosity and the broad H$\beta$ line, which implied a deviation from simple theoretical models predicting a slope of $0.5$. The relation was further modified for lower-redshift sources using the H$\beta$ RM and the luminosity at 5100 $\angstrom$ \citep{2006ApJ...644..133B,bentz2009a,bentz2013} taking into account the host galaxy starlight. After the removal of the host-galaxy starlight from the total luminosity, the power-law slope was determined to be $\alpha=0.533^{+0.035}_{-0.033}$ \citep{bentz2013}, which is consistent with the theoretical expectation from simple photoionization models. The slope of $\alpha=1/2$ simply follows from the ionization parameter of a BLR cloud, $U=Q(H)/(4\pi R^2 c n_{\rm H})$, where $R$ is the cloud distance from an ionizing continuum, $c$ is the speed of light, $n_{\rm H}$ is the number density, and $Q(H)=\int_{\nu_i}^{\infty}L_{\nu}/h\nu \, \mathrm{d}\nu$ is the ionizing photon flux. In general, one can assume that the ionization conditions as well as the gas density in the BLR are comparable for all AGN, i.e. the product $Un_{\rm H}$ is constant, from which follows $R_{\rm BLR}\propto Q(H)^{1/2} \propto L^{1/2}$ \citep{2013peag.book.....N}.

\begin{figure*} 
\centering
\includegraphics[width=0.65\textwidth]{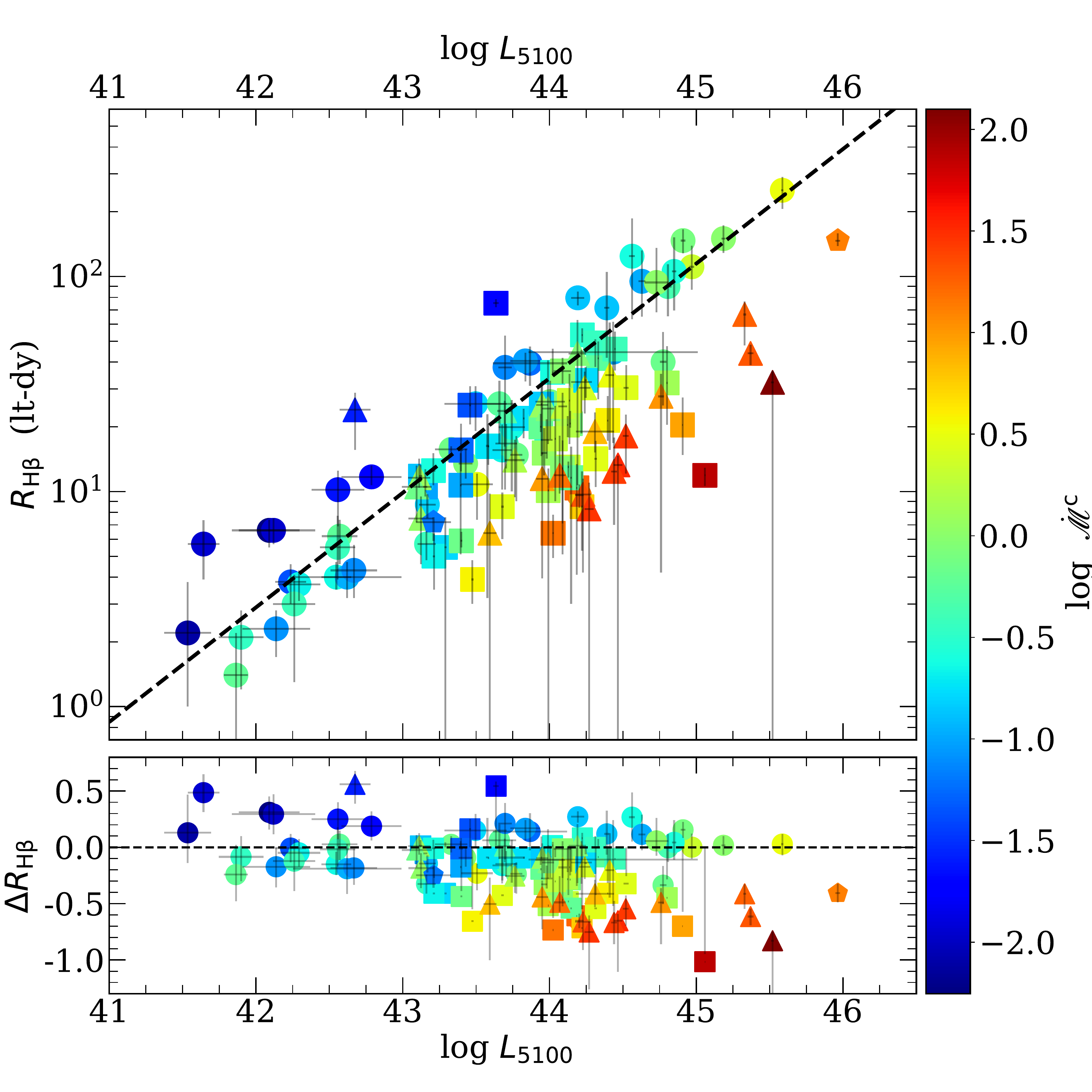}
\caption{TOP PANEL: $R_{H\beta}-L_{5100}$ relation for SEAMBH (triangles), SDSS-RM (squares), Bentz Collection (circles), NGC5548 and 3C 273 (pentagons). BOTTOM PANEL: Difference between the BLR size estimated from the observed time delay and the predicted ones by the RL relation (black line,  Eq.~\ref{eq_bh_mass}). In both panels, colors indicate the variation in dimensionless accretion rate, \mdotc.\label{fig:RL}}
\end{figure*}

In Figure \ref{fig:RL} the \RL\ relation created from 117 \hb\ reverberation-mapped AGN. The full sample consists of 48 sources previously monitored by \citet{bentz2009a, bentz2014, barth2013, pei2014, bentz2016a}, and \citet{fausnaugh2017}, 25  super-Eddington sources of the SEAMBH project  \citep[Super-Eddington Accreting Massive Black Holes,][]{Du2015, Wang2014, Hu2015, Du2015, Du2016, Du2018}, 44 sources from the SDSS-RM \citep{2017ApJ...851...21G} sample, and the recent monitoring for NGC5548 \citep{lu2016} and 3C273 \citep{zhang2018}. See more details concerning the sample in \citet{martinezaldama2019}.

The importance of the \RL\, relationship lies in the fact that it is the basis of secondary black-hole mass estimates, especially for high-$z$ AGN \citep{1998ApJ...505L..83L,1999ApJ...526..579W,2002MNRAS.337..109M,2006ApJ...641..689V}. In principle, by knowing the optical luminosity of the source, one can infer the black-hole mass from a single-epoch spectrum (by extracting the broad-line FWHM), which makes it especially useful for large statistical surveys of sources throughout the cosmic history. Given the importance of \RL\,relation, it is necessary to verify it using different emission lines from the LIL BLR region and higher-redshift sources, which we will analyze in subsequent section for the case of MgII line and the redshift of 0.9 (a complete and updated MgII based RM sample is reported in Zaja\v{c}ek et al. 2019, submitted to ApJ). 

RM of quasars, using both the BLR reverberation technique and the accretion disc radiation reprocessing, was suggested to be utilized in cosmological studies \citep{1999MNRAS.302L..24C,2002ApJ...581L..67E,2003MNRAS.339..367H}. In particular, the \RL\, relation directly enables to turn reverberation-mapped quasars into ``standardizable'' luminosity candles \citep{2011A&A...535A..73H,2011ApJ...740L..49W,2013A&A...556A..97C,bentz2013}. Under the assumption that the \RL\, power-law relationship applies not only to H$\beta$ line but also other LIL lines, one could apply it to intermediate- and high-redshift quasars whose co-moving time-delay was determined to infer their absolute luminosities. From measured monochromatic flux densities of the sources, one can infer the luminosity distance $D_{\rm L}=\sqrt{L_{5100}/4\pi F}$ and the Hubble diagram of reverberation-mapped quasars. The critical point here is to reduce the scatter along the radius-luminosity relation, both by improving the RM time-lag precision and by correctly subtracting the host starlight \citep{2011A&A...535A..73H}.  

The paper is structured as follows. In Section~\ref{sec_radius_luminosity}, we report on the detection of MgII line time-delay for the intermediate-redshift quasar CTS C30.10 and its consistency with the radius-luminosity relation previously derived for H$\beta$ emission in lower redshift sources. We continue with Section~\ref{sec_accretion_RL} to analyze in detail how accretion rate affects the position of sources along the \RL\, relation. We summarize the initial cosmological constraints as derived from the current reverberation-mapped quasar sample in Section~\ref{sec_cosmology}. First steps of modelling stochastic lightcurves in the context of reverberation mapping with the \textit{Large Synoptic Survey Telescope} (LSST) are analyzed highlighting a novel filtering algorithm to select higher quality photometric data in Section~\ref{sec_LSST}. We discuss a few open problems in Section~\ref{section_discussion} and finally conclude with Section~\ref{section_conclusions}.

\section{Radius-luminosity relationship towards higher redshifts:time-lag determination of  M\lowercase{g}II response in CTS C30.10}
\label{sec_radius_luminosity}

Previously the power-law radius-luminosity relation, $R_{H\beta}\propto L_{5100}^{\alpha}$, was observationally constrained using H$\beta$ broad line for low-redshift sources up to $z =0.292$ \citep{2000ApJ...533..631K,bentz2009a,bentz2013}, with the scatter as low as $0.13$ dex and the power-law index of $\alpha=0.546^{+0.027}_{-0.028}$ \citep{bentz2013} after the host starlight correction. The analysis using H$\beta$ broad line limits the redshift of the sources up to $z\sim 0.6$ for optical observations with the maximum wavelength limit of 8000\angstrom. However, other LIL broad lines can extend the redshift range to intermediate values. MgII line with the rest wavelength $\sim$ 2798 \AA~ is suitable for the reverberation-mapping studies in the redshift range $z=0.4-1.5$ for ground based telescopes, which is also utilized for the sample of seven objects monitored by the South African Large Telescope (SALT) telescope \citep{2013A&A...556A..97C}.

Between December 6, 2012 and December 10, 2018, we monitored the quasar CTS C30.10 using the $10$-meter SALT telescope in order to detect the time-lag of the LIL emission-line of MgII with respect to the quasar continuum. The hypothesis was to probe the consistency of the radius-monochromatic luminosity relation towards higher redshifts and higher absolute luminosities. The quasar CTS C30.10 was discovered as a bright source in the Calan-Tololo survey \citep{1993RMxAA..25...51M} among 200 newly discovered quasars with the visual magnitude of $V=17.2$ (NED) at the intermediate redshift of $z= 0.90052$ \citep{2014A&A...570A..53M}. The equatorial coordinates of the source are $RA= 04h47m19.9s$ and $Dec=-45d37m38s$ (J2000.0).

The MgII emission-line light curve was constructed based on the long-slit spectroscopy mode of the SALT telescope with the slit width of 2'' \citep[for details, see ][]{2014A&A...570A..53M, 2019ApJ...880...46C}. The MgII line was extracted from the spectral energy distribution of CTS C30.10 in the wavelength range $2700-2900\,\angstrom$ by subtracting the two spectral components, namely the power-law component due to the accretion disc thermal emission, and FeII - line pseudo-continuum. The MgII broad line was found to consist of two kinematic components -- redshifted and blueshifted Lorentzian profiles\footnote{the use of the double Lorentzian profiles is motivated from the study for this source in \citet{2014A&A...570A..53M}. The value of the $\chi^2$ obtained for the double Lorentzian is the lowest (see their Table 3).} -- each of which further consists of a doublet at $2796.35\, \angstrom$ and $2803.53\,\angstrom$ with the doublet ratio of $1.6$. The equivalent width (EW) of the MgII line, EW$\sim F_{\rm line}/F_{\rm cont}$, was calculated by numerical integration of all four spectral components -- 2 kinematic and 2 doublet components. Because of the two kinematic components of MgII line, CTS C30.10 belongs to type B quasars \citep{2007ApJ...666..757S, 2014A&A...570A..53M} characterized by lower Eddington ratios in the range $\log{(L/L_{\rm Edd})}=0.01-0.2$ \citep{2011BaltA..20..427S}, while type A sources exhibit a single component MgII line of a Lorentzian shape \citep{1997ApJ...489..656L,2001A&A...372..730V,2002ApJ...566L..71S,2010MNRAS.403.1759Z,2009NewAR..53..198S,2011BaltA..20..427S,2012ApJS..202...10S}. The origin of two components is still quite uncertain although the presence of the second, asymmetric component and the overall FWHM is consistent with our source belonging to Pop. B. It could either imply the presence of second emission region due to absorption or scattering or it could hint at the origin of LIL lines close to the disc plane, resembling thus the disc kinematics, which would naturally lead to two components when viewed off-axis. 

The continuum light curve was obtained from the OGLE-IV survey in the $V$-band and SALTICAM in the $g$-band. The SALTICAM observations were shifted freely to match the overlapping V-band values. Since the SALT observations are not spectrophotometric, MgII flux density was obtained using its EW and the continuum flux density.

The normalized continuum dispersion was $6.0\%$, while the line dispersion was $5.2\%$, which is slightly lower but comparable within uncertainties. This is consistent with the simple reprocessing scenario, where the central disc provides all the ionizing UV photons which are absorbed and scattered by BLR clouds at larger distances. This allowed us to use both the continuum and the line-emission light curves to infer the time-delay $\tau_{\rm MgII}$ of the MgII broad line.

\subsection{Time-lag determination of MgII line using cross-correlation function}
\label{subsec_time_lag}

First, we used the standard interpolated cross-correlation function (ICCF) to infer the time-delay between the MgII line-emission and the V-band continuum, which corresponds to the continuum around the redshifted MgII line. The ICCT requires regularly sampled datasets, while realistic light curves are unevenly sampled. The regular timestep is achieved by interpolating the continuum light curve to time-shifted emission-line light curve or vice versa. Typically, both interpolations are averaged to obtain the symmetric ICCF. Given the two light curves $x_{i}$ and $y_{i}$ sampled at discrete time intervals $t_{i}$ $(i=1,...,N)$ with the regular time-step $\Delta t=t_{i+1}-t_{i}$, we can define the cross-correlation function (CCF) as,

\begin{equation}
    CCF(\tau_k)=\frac{(1/N)\sum_{i=1}^{N-k}(x_i-\overline{x})(y_{i+k}-\overline{y})}{[(1/N)\sum_{i=1}^{N}(x_i-\overline{x})^2]^{1/2}[(1/N)\sum_{i=1}^{N}(y_i-\overline{y})^2]^{1/2}}\,,
    \label{eq_CCF}
\end{equation}
where $\overline{x}$ and $\overline{y}$ are light curve mean values, respectively, and $\tau_{k}=k \Delta t$ ($k=0,...,N-1$) is the time-shift of the second light curve with respect to the first one at which the CCF is evaluated.  

\begin{figure}
    \centering
    \includegraphics[width=\textwidth]{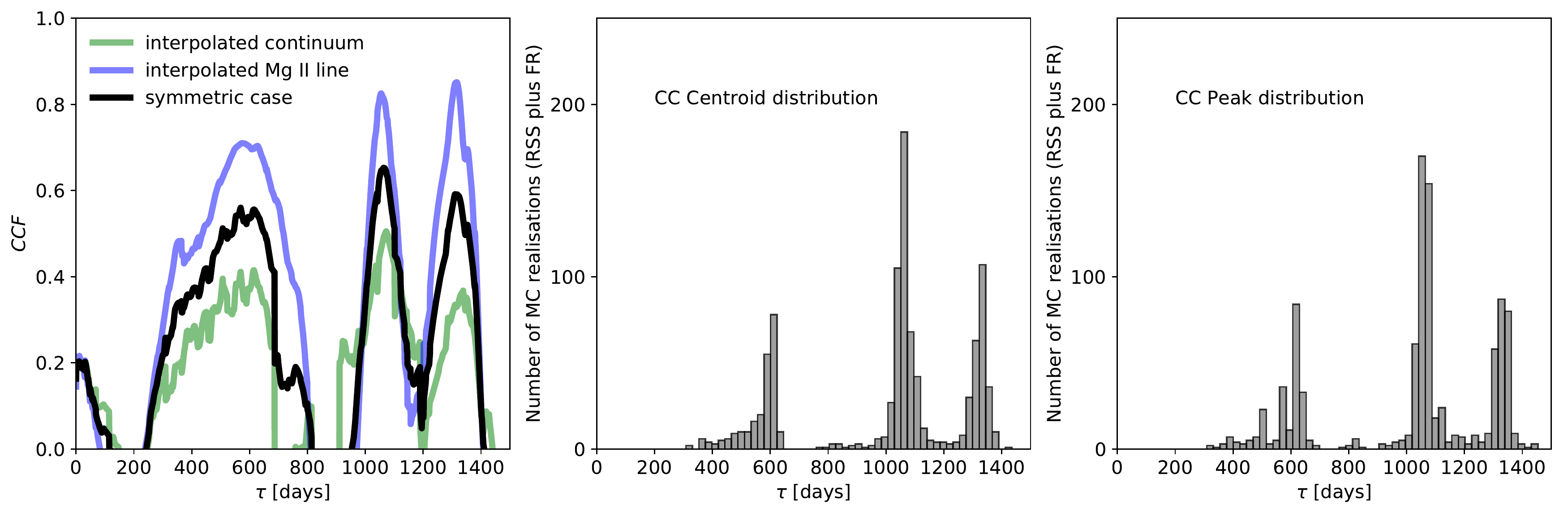}
    \caption{Results of the interpolated cross-correlation analysis between the continuum and MgII line emission. \textbf{Left panel:} The CCF value as a function of the time-delay for the continuum light curve interpolated to the emission-line light curve (green line), for the emission-line light curve interpolated to the continuum light curve (blue line), and the symmetric interpolation (black line). \textbf{Middle panel:} Cross-correlation centroid distribution. \textbf{Right panel:} Cross-correlation peak distribution.}
    \label{fig_iccf}
\end{figure}

We make use of the \textmyfont{PYTHON} code \textmyfont{PYCCF} \citep{2018ascl.soft05032S}, which is the implementation of earlier CCF analysis by \citet{1987ApJS...65....1G} and \citet{1998PASP..110..660P}. We calculate CCF separately for interpolated continuum, interpolated emission-line curves as well as the symmetric case. Using thousand Monte Carlo runs of the combined random subset selection (RSS) and flux randomization (FR), we obtain distributions of the CCF centroid (CCFC) and the CCF peak (CCFP), from which we can calculate the corresponding uncertainties, see Fig.~\ref{fig_iccf}. 

\begin{table}[tbh]
\centering
\caption{Summary of methods used for calculating the time-delays for the quasar and their uncertainties expressed with respect to the observer's frame.}
\begin{tabular}{c|c|c}
\hline
\hline
Method & Note & Time-delay [days]\\
\hline 
ICCF & Symmetric interpolation - Centroid    &   $1060^{+262}_{-457}$\\
ICCF & Symmetric interpolation - Peak        &   $1064^{+260}_{-446}$\\
ICCF & Interpolated continuum - Centroid     &   $1053^{+219}_{-467}$\\
ICCF & Interpolated continuum - Peak         &   $1060^{+201}_{-491}$\\
ICCF & Interpolated line emission - Centroid &   $1228^{+84}_{-319}$\\
ICCF & Interpolated line emission - Peak    &   $1288^{+39}_{-310}$\\
DCF & time-step of 100 days                  &   $1030^{+106}_{-140}$\\
zDCF &  -                                    &   $1050^{+32}_{-27}$\\
\hline
\end{tabular}
\label{table_time_delay_method}
\end{table}

In Table~\ref{table_time_delay_method}, we summarize the time-delay results for all considered cases -- the time-delay values are expressed with respect to the observer's frame ($\tau_{\rm{observer}} = (1+z)\times \tau_{\rm{rest}}$). For the symmetric interpolation, the centroid time-delay is $\tau_{\rm ICCF}=1060^{+262}_{-457}$ days. In general, the cross-correlation centroid and peak distributions have three peaks, see Fig.~\ref{fig_iccf}, while the peak at $\sim 1060$ is the most prominent. At this value, ICCF also reaches the largest value for the symmetric and interpolated-continuum cases, see Fig.~\ref{fig_iccf} (left panel), while the interpolated-line case peaks at larger time-delays. This offset may be due to the fact that the line-emission dataset is not so densely covered as the continuum light curve, which leads to artefacts when interpolating it to denser continuum light curve.

Second, we applied the discrete correlation function (DCF), which does not interpolate between the light curves and is thus better suited for unevenly sampled datasets with known measurement errors. The DCF was described by \citet{1988ApJ...333..646E} and applied routinely to search for light curve correlations and time-lags. The first step in the DCF analysis is to look for data pairs $(x_i,y_j)$ that fall into the time-delay bin of $\tau-\delta \tau/2 \leq \delta t_{ji} \leq \tau+\delta \tau/2$, where $\tau$ is the given time-delay, $\delta \tau$ is the time-delay bin and $\delta t_{ji}=t_j-t_i$. Given $M$ such data pairs, one calculates a corresponding number of unbinned discrete correlation coefficients $UDCF_{ij}$ in the following way,
\begin{equation}
    UDCF_{ij}=\frac{(x_i-\overline{x})(y_j-\overline{y})}{\sqrt{(s_{x}-\overline{\sigma}^2_{x})(s_{y}-\overline{\sigma}^2_{y})}}\,,
    \label{eq_udcf}
\end{equation}
where $\overline{x}, \overline{y}$ are the light curve means for a given time-delay bin. Other parameters $s_{x}$ and $s_{y}$ stand for the variances, and $\overline{\sigma}^2_{x}$ and $\overline{\sigma}^2_{y}$ are the mean measurement errors for a given time-delay bin. Subsequently, the DCF coefficient can be calculated for a given time-lag bin by averaging in total $M$ $UDCF_{ij}^{M}$ values,
\begin{equation}
    DCF(\tau)=\frac{1}{M}\sum_{M} UDCF_{ij}^{M}\,.
    \label{eq_dcf}
\end{equation}
where, \textit{M} is the number of light curve pairs that fall into a particular time-lag bin. The uncertainty can be estimated by the relation,
\begin{equation}
    \sigma_{\rm DCF}(\tau)=\frac{1}{M-1}\sqrt{\sum_{M} [UDCF_{ij}^{M}-DCF(\tau)]^2}\,.
    \label{eq_sigma_DCF}
\end{equation}

We applied the \textmyfont{PYTHON} code of Damien Robertson \citep[see][for the code implementation]{2015MNRAS.453.3455R}, which calculates the DCF with the possibility of Gaussian weighting for the matching pairs of light curve points. The code allows us to select a different size for equal time-bins as well as the searched interval for the time-delay. We expand the possibilities of the code by adding the bootstrap technique to assess the significance of individual DCF peaks and to better estimate their uncertainties.  

\begin{figure}
    \centering
    \includegraphics[width=0.48\textwidth]{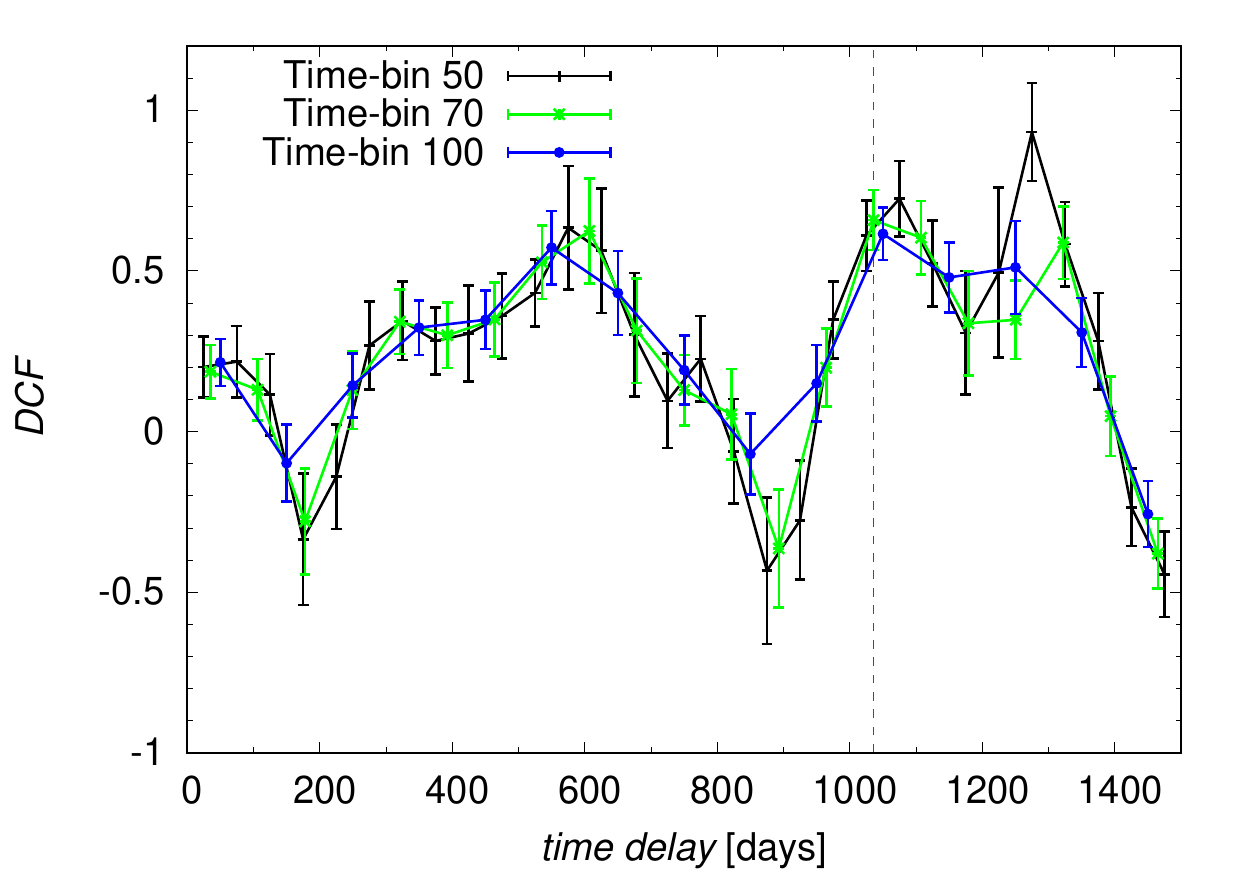}
    \includegraphics[width=0.48\textwidth]{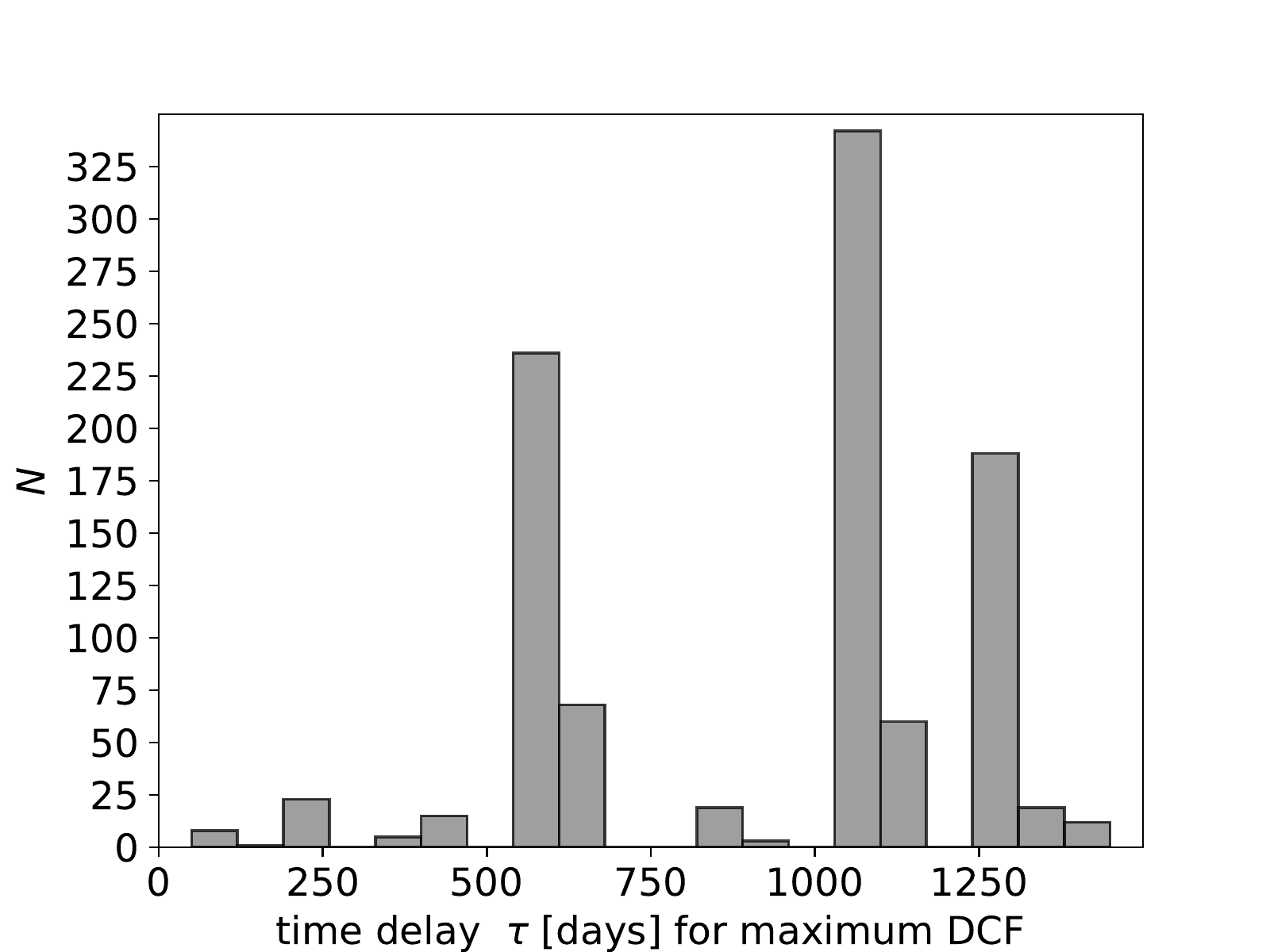}
    \caption{Results of the Discrete Correlation Function (DCF). \textbf{Left panel:} DCF as a function of time-delay in the observer frame for three different sizes of the time-bin: 50, 70, and 100 days, see the legend. The largest DCF is consistently at the time-delay of 1050 days for bigger time-steps of 70-100 days. \textbf{Right panel:} Histogram of 1000 bootstrap runs using random subsample selection. The largest peak is at $\tau_{\rm DCF}=1030^{+106}_{-140}$ days.}
    \label{fig_DCF}
\end{figure}

The DCF uses equal time-step binning and it is thus quite sensitive with respect to the time-bin size. We test this using three different timesteps -- 50, 70, and 100 days -- which leads to the decrease in the mean DCF values for the time-delay at $\sim 1275$ days, which is the most prominent for the smaller time-bins but significantly gets smaller for larger time-bins, see Fig.~\ref{fig_DCF} (left panel). For the time-bin size of 100 days, the largest DCF of $0.62$ is for 1050 days, followed by the peak at 550 days. Moreover, we verify the significance of individual peaks by running 1000 bootstrap realizations where we randomly create subsamples from both light curves simultaneously. We obtain the distribution of time-delay peaks (for which the DCF value is the largest for each run) in Fig.~\ref{fig_DCF} (right panel), from which we obtain the peak time-delay with upper and lower 1$\sigma$ uncertainties, $\tau_{\rm DCF}=1030^{+106}_{-140}$ days, which we also include in Table~\ref{table_time_delay_method}. 

\begin{figure}[tbh]
    \centering
    \includegraphics[width=\textwidth]{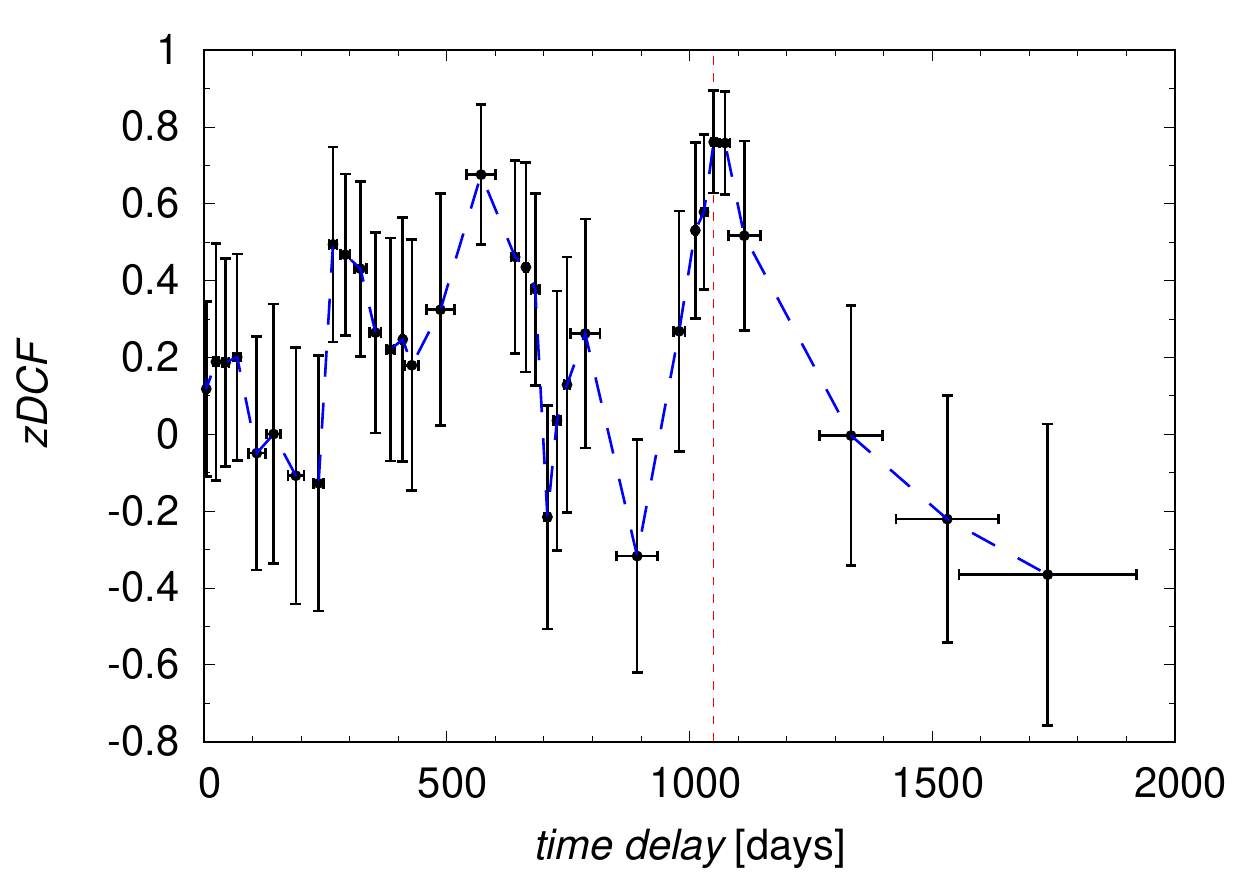}
    \caption{Results of the $z$-transformed discrete correlation function (zDCF). We show the zDCF as a function of the time-delay in the observer frame. The red vertical dashed line marks the time-delay with the largest value of a zDCF coefficient.}
    \label{fig_zdcf}
\end{figure}

Finally, several biases and problems of the cross-correlation function which primarily stand out when dealing with sparse and heterogeneous light curves were taken into account in $z$-transformed DCF \citep{1997ASSL..218..163A}. In comparison with the discrete correlation function \citep{1988ApJ...333..646E}, which bins data pairs into equal time-bins, $z$-transformed DCF applies equal population binning. It is, therefore, a more suitable and robust method for sparse, irregularly sampled, and heterogeneous pairs of light curves with as few as 12 points per population bin.

We show the calculated zDCF as a function of time delay in the observer frame in Fig.~\ref{fig_zdcf}. The largest zDCF is for the time-delay at $\tau=1050$ days (zDCF$=0.76$), followed by the smaller peak at $\tau=571$ days (zDCF$=0.68$). We verify the significance of the peak using the maximum likelihood, which also enables us to estimate its uncertainty. We obtain the peak value of $\tau_{\rm zDCF}=1050^{+32}_{-27}$ days, which is also listed among other methods in Table~\ref{table_time_delay_method}.

We summarize the time-delay value for CTS C30.10 by calculating the average value for the most prominent peak at $\sim 1050$ days in the observer frame using Table~\ref{table_time_delay_method}. We get the mean value of $\overline{\tau}_{\rm obs}=1048^{+94}_{-156}$ days in the observer frame, which translates into $\overline{\tau}_{\rm source}=551^{+49}_{-82}$ days in the source frame. 

\subsection{Radius-luminosity relation for MgII}

As for H$\beta$ broad line, we construct the radius-luminosity relation for MgII line taking into account our detection of time-lag for MgII in CTS C30.10 as well as the measurements of other sources that also have RM data of MgII: 6 sources from \citet{2016ApJ...818...30S}, CTS252 from \citet{2018ApJ...865...56L}, and NGC4151 from \citet{2006ApJ...647..901M}. For an overview of the sources and their characteristics, see Table 3 in \citet{2019ApJ...880...46C}.

\begin{figure}[tbh]
    \centering
    \includegraphics[width=\textwidth]{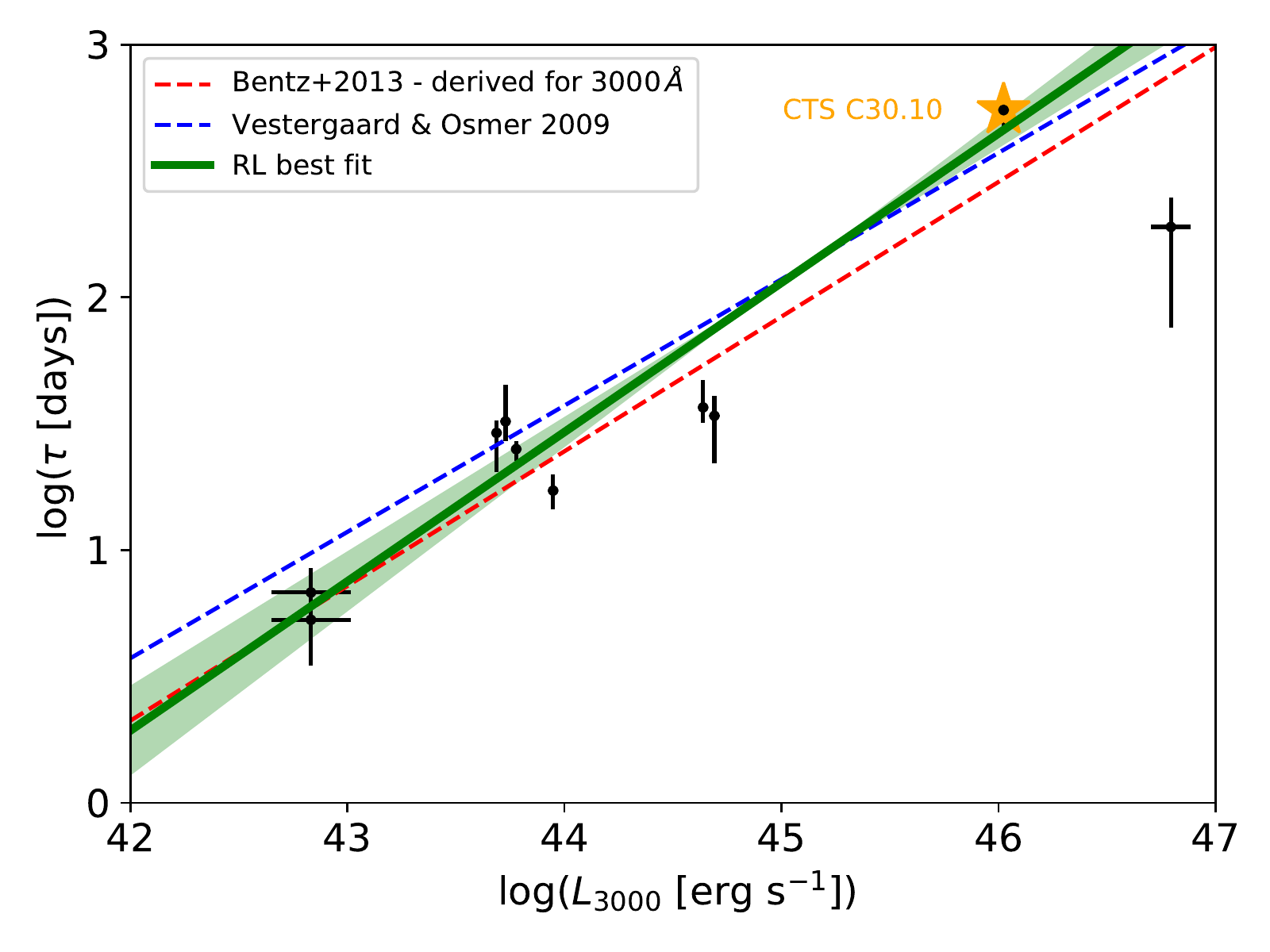}
    \caption{The radius-luminosity relation for MgII broad line using the sources as listed in Table 3 in \citet{2019ApJ...880...46C} and the new measurement of CTS C30.10 (orange star). The red dashed line stands for the standard Bentz relation \RL\, \citep{bentz2013}, derived for $3000\,\angstrom$ using the bolometric corrections from \citet{2019MNRAS.488.5185N}: $\log(R_{\rm BLR}/1 \text{lt-day})=1.391+0.533\log{(L_{3000}/10^{44}\,{\rm erg\,s^{-1}})}$. The blue dashed line is the MgII-based radius-luminosity relation derived by \citet{2009ApJ...699..800V}: $\log(R_{\rm BLR}/1 \text{lt-day})=1.572+0.5\log{(L_{3000}/10^{44}\,{\rm erg\,s^{-1}})}$. The green solid line is the best-fit result using the general prescription $\log(R_{\rm BLR}/1 \text{lt-day})=K+\alpha\log{(\lambda L_{\lambda}/10^{44}\,{\rm erg\,s^{-1}})}$ with $K=1.47 \pm 0.06$ and $\alpha=0.59\pm 0.06$. The shaded region corresponds to 1$\sigma$ uncertainties.}
    \label{fig_RL_MgII}
\end{figure}

In Fig.~\ref{fig_RL_MgII}, we show the radius-luminosity plot using MgII line time-delay in the rest frame and the monochromatic luminosity at 3000 \angstrom. We compare the localization of sources with the Bentz \RL\, relation \citep{bentz2013}, $\log(R_{\rm BLR}/1 \text{lt-day})=1.391+0.533\log{(L_{3000}/10^{44}\,{\rm erg\,s^{-1}})}$, where the monochromatic luminosity at 5100 \angstrom\, is related to the luminosity at 3000 \angstrom\, by an approximate scaling $L_{5100}=(5/8)^{5/4} L_{3000}$ as inferred from the luminosity-dependent bolometric corrections by \citet{2019MNRAS.488.5185N}. For comparison, we also display the MgII-based radius-luminosity relation as derived by \citet{2009ApJ...699..800V}: $\log(R_{\rm BLR}/1\text{lt-day})=1.572+0.5\log{(L_{3000}/10^{44}\,{\rm erg\,s^{-1}})}$. We see that the majority of the sources follow radius-luminosity relation within uncertainties, with CTS252 being an outlier, which could be the hint of the trend of a smaller time-delay for higher Eddington-ratio sources as was already studied for H$\beta$ measurements \citep{Du2018, martinezaldama2019}. Based on the smaller MgII line width, CTS252 is a higher Eddington-ratio source in comparison with CTS C30.10 \citep{2019ApJ...880...46C}, but the trend needs to be confirmed for a higher number of MgII measurements.

Excluding CTS252, we fit the radius-luminosity dataset with the general relation $\log(R_{\rm BLR}/1 \text{lt-day})=K+\alpha\log{(\lambda L_{\lambda}/10^{44}\,{\rm erg\,s^{-1}})}$. We obtain the best-fit parameters of $K=1.47 \pm 0.06$ and $\alpha=0.59\pm 0.06$. In Fig.~\ref{fig_RL_MgII}, we see that the best-fit radius-luminosity relation passes nearly through the CTS C30.10 point in comparison with the Bentz relation, which is due to a slighly higher mean slope --  $\alpha=0.59$ instead of $\alpha=0.53$. However, the slopes do not differ within uncertainties and hence the radius-luminosity relation with the expected scaling of $R\sim L^{1/2}$ so far holds for MgII time-lag measurements.

\section{Accretion rate effect over the Radius-Luminosity relation}
\label{sec_accretion_RL}

The \RL\ relation offers the possibility of estimate the luminosity distance (\DL) independently of the redshift, which is suitable for cosmological applications \citep[][and references therein]{martinezaldama2019}. Also, it shows a low scatter \citep[0.13dex,][]{bentz2013} and although objects, like NGC5548, show a large variation, the scatter is  within uncertainties. However, the recent inclusion of AGN radiating close to the Eddington limit (super-Eddington sources) has increased the scatter significantly. This kind of sources are located on the right-bottom side of the relation (See Figure~\ref{fig:RL}), which implies that the time delay of super-Eddington sources is shorter than the predicted by the \RL\ relation. 

Super-Eddington sources tend to show an extreme behavior respect to the general AGN population: large densities ($n_{H}\sim10^{13}$  cm$^{-3}$), low-ionization parameters (log $U<-2$), high intensity of lowest ionization emission lines like FeII and strong outflows in the high-ionization like CIV$\lambda1549$ (\citealt{2019A&A...627A..88M} and references therein). These features are probably explained by an optically and geometrically thick slim disk \citep{wangshielding2014}, which shields the line emitting region gas from the most intense UV radiation \citep{marziani2018}. This peculiar behavior is also reflected in the \RL\ relation by the super-Eddington sources of SEAMBH sample, but also by some objects of SDSS-RM sample. One-third of the SDSS-RM shows an Eddington ratio higher than the average Eddington ratio in Bentz sample, then they also a departure from the expected \RL\ relation. Therefore, a correction is needed not just for super-Eddington sources, but all sources have to be rescaled.

To estimate this correction to the measured time delay as a function of the accretion rate, we estimate the black hole mass (Equation~\ref{eq_bh_mass}) considering a virial factor anti-correlated with the FWHM of the line recently proposed by \citet{mejia-restrepo2018}, which in some sense corrects for the orientation effect. For the accretion rate, we will consider the dimensionless accretion rate (\mdotc) introduced by \citet{Du2016}. Details of the estimated values are reported in \citet{martinezaldama2019}. To estimate the departure from the \RL, we consider the parameter $\Delta R_{\mathrm{H\beta}}=\mathrm{log}\,\left( \frac{\tau\mathrm{_{obs}}}{\tau\mathrm{_{{H\beta_{\,R-L}}}}}\right)$, which is the difference between the time delay observed and estimated from the \RL\ relation. In the bottom panel of Fig.~\ref{fig:RL} is shown the behavior of \DRhb\ as a function of $L_{51  00}$. Also, in colors, it is shown the variation of the dimensionless accretion rate, where it is clearly observed that largest departures correspond to  highest \mdotc\ values. The relation between \DRhb\ and \mdotc\ can be described by the linear relation:
\begin{equation}\label{equ:DRHb}
\Delta R\mathrm{_{H\beta, \dot{{M}}\mathrm{^{c}}}} = \left(-0.283\pm0.017\right) \mathrm{log}\dot{{M}}^\mathrm{{c}}+\left(-0.228\pm0.016\right).
\end{equation}
With this relation, the observed time delay can be corrected by the dimensionless accretion rate effect, decrease the scatter and recover the time delay predicted by the \RL\ relation. However, this relation is strongly dependent on the virial factor selected for the black hole mass and the accretion rate estimations. The virial factor is still an open problem and although many formalisms have been proposed anyone can be applied to the general AGN population. \fblrc\ was modeling with a scarcity of narrow profiles (FWHM$<$2000\kms), which represent 30$\%$ of our full sample. The inclusion of narrow profiles in its modeling could modify the exponent of the anti-correlation and modify our results. 


\section{Cosmology with reverberation-mapped sources}
\label{sec_cosmology}

\begin{figure*} 
\centering
\includegraphics[width=0.60\textwidth]{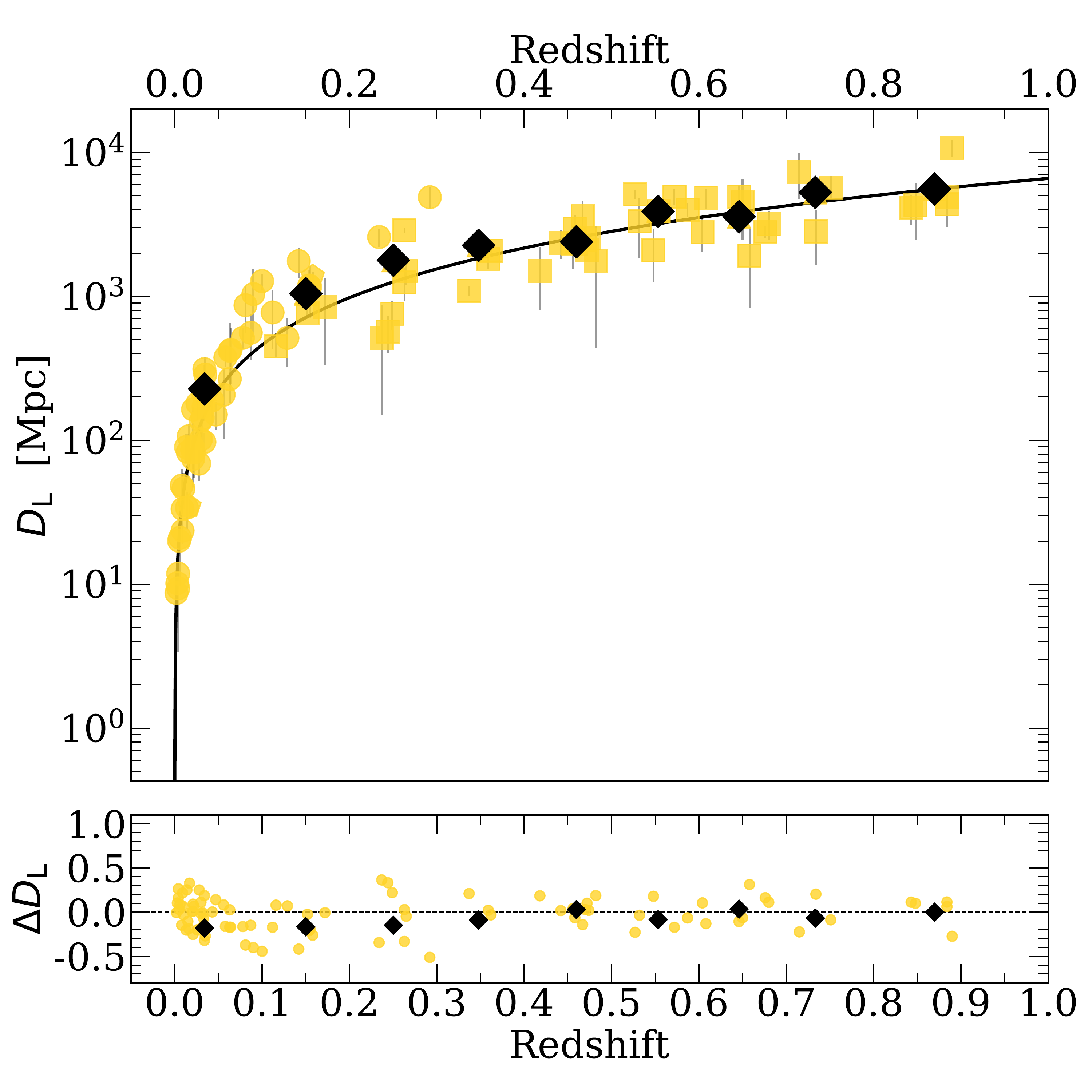}
\caption{Hubble diagram after the correction by dimensionless accretion rate. Markers and colors are the same as in Figure~\ref{fig:RL}. The black lines indicates the expected luminosity distance based on the standard $\Lambda$CDM model. The black diamonds represent the average values for the \LD{} considering redshift bins of $\Delta z=0.1$. The bottom panel shows the difference between the expected luminosity distance and the observed one.} \label{fig:cosmo}
\end{figure*}

Recovering the low scatter of the \RL\ relation after the correction by the dimensionless accretion rate, we were able to build a Hubble diagram (Fig.~\ref{fig:cosmo}) which relates the distance to the sources with the redshift ($z$) or the velocity recession. The slope of the relation between these two parameters is the value of the Hubble constant  ($H_0$), which determines the current expansion rate of the Universe. This parameter has been estimated in the early Universe ($z<1000$) using the Cosmic Microwave Background (CMB, \citealt{planck2018}) and in the late Universe (up to z $\sim$ 1.5) has been using Cepheid stars (\citealt{2019MNRAS.484L..64S} and references therein) and Supernovae Ia (SNIa, \citealt{2018ApJ...869...56B, 2019MNRAS.486.2184M, riess2019}). The most recent results from the \citet{planck2018} indicate a value of $H_0=$67.66$\pm0.42$ \kms\ Mpc$^{-1}$. While the recent ones employing observations of Cepheids stars and SNIa from the Hubble telescope give a value of 74.3$\pm$1.42 \kms\ Mpc$^{-1}$ \citep{riess2019}. The precision in the determination, less than $2\%$, discards the possibility of an error measurement, therefore the disagreement suggests a change in the Hubble constant in the different epochs of the Universe and new physics is required to solve the problem. This problem is called Hubble constant tension. The recent results of \citet{risaliti2019} using quasars at $z\sim4$ also show a small gap between the standard $\Lambda$CDM model and the fit performed. The large redshift range covered by quasar ($0<z<7$) is suitable for estimating the Hubble constant and addressing the Hubble constant tension. 

To get the luminosity distance (\LD), we estimate $L_{5100}$ from the Equation~\ref{eq_bh_mass} and after we use the relation: $D\mathrm{_L}= \left(\frac{ L_{5100}}{ 4\,\pi\,F_{5100} }\right)^{1/2}$. In the Figure~\ref{fig:cosmo}  the black line marks the expected model according to the $\Lambda$CDM model. As a visual representation, we show the \LD\ considering redshift bins of $\Delta z=0.1$ (black symbols), which do not have any statistical significance by the number of points. The standard deviation of the errors (0.31, shown in the bottom panel) does not show any particular trend but the dispersion is still large in comparison with other, more matured methods.  For the determination of the cosmological constant, we assumed the $\Lambda$CDM model and a value for the Hubble constant of $H_0=$67.66$\pm0.42$ \kms\ Mpc$^{-1}$. Looking for the best fit using a minimization method ($\chi^2$), we get the best values for $\Omega_m$ and $\Omega_\Lambda$. Our results are in agreement with the standard model within 2$\sigma$ confidence level, however, it is not yet ready to provide new for cosmological constraints, considering that error associated with CMB, Cepheids stars and SNIa are less than 2$\%$. 

There are still some systematic errors related that have to be corrected. In the previous section, we proposed a correction by the accretion rate effect, but this correction has to be improved since the virial factor has associated large uncertainties. Then, we have to work for clarifying the dynamics of the BLR and the orientation effect. Another important source of error is the method employed in the determination of the time delay. The delay measurements used here are a compilation of the results from the literature, and each group uses different criteria and methods, which also introduces an error when different samples are compared.

\begin{figure}[tbh]
    \centering
    \includegraphics[width=0.48\textwidth]{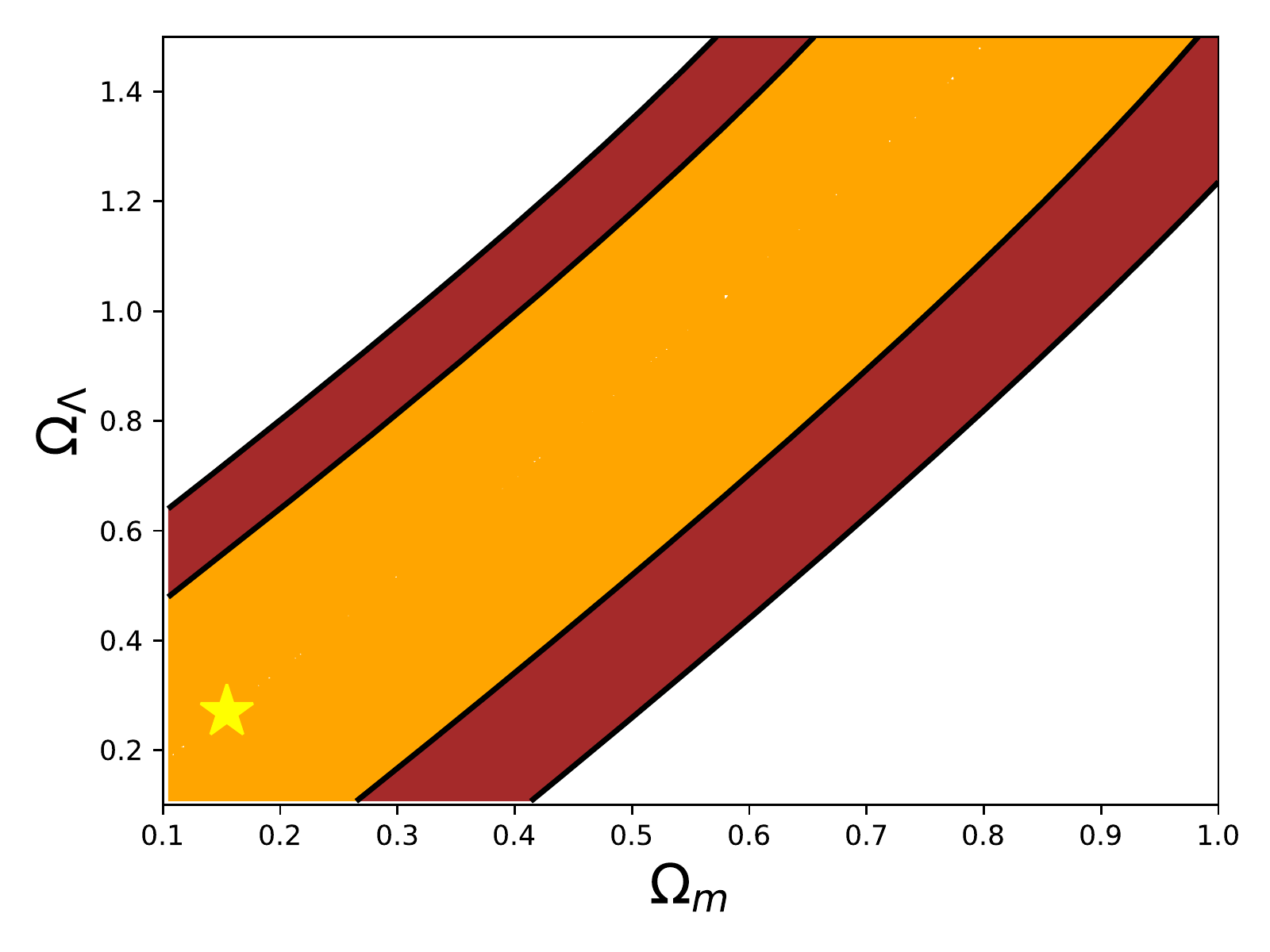}
    \includegraphics[width=0.48\textwidth]{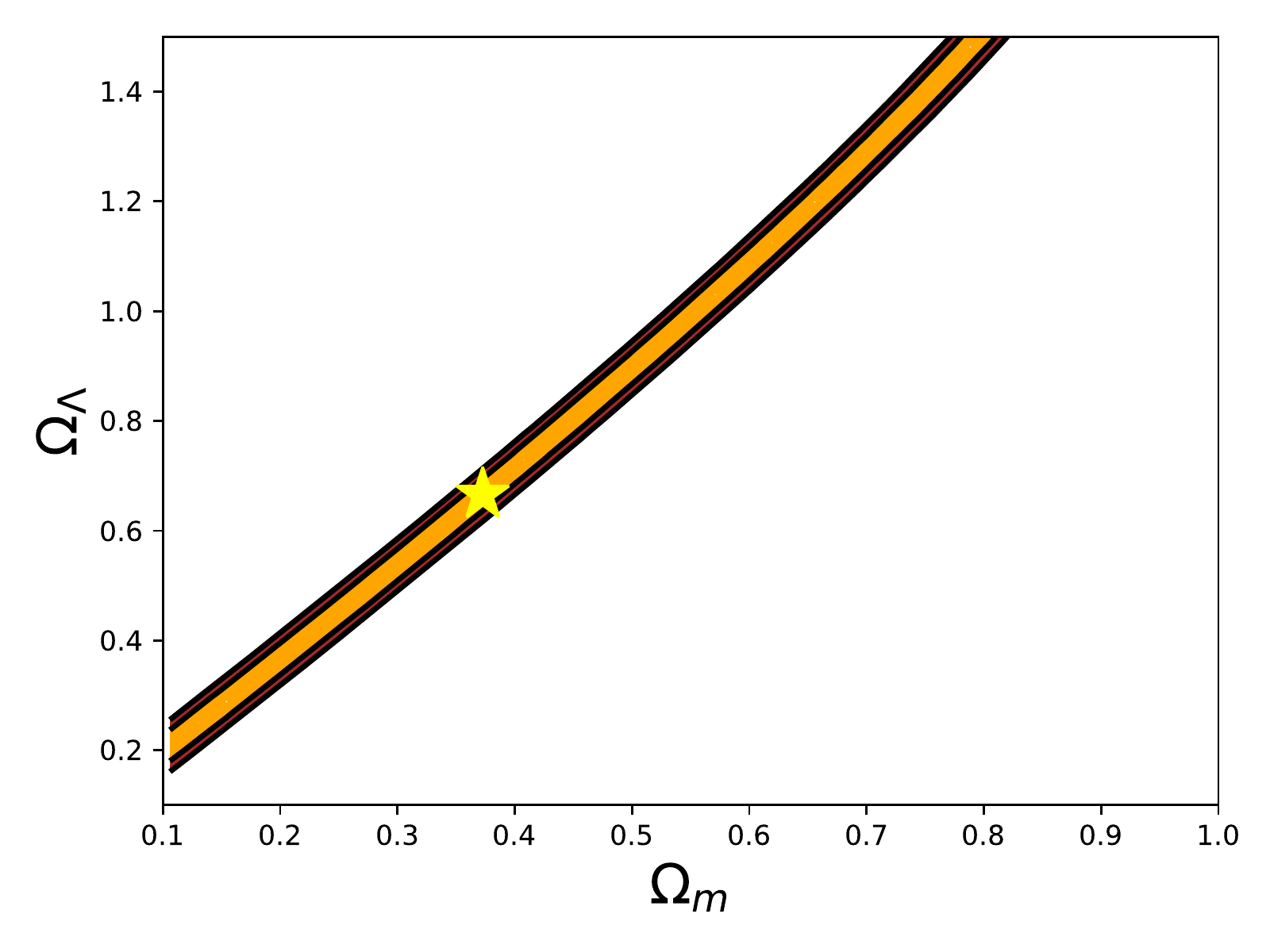}
    \caption{Effect of the time-delay uncertainty on constraining cosmological parameters for the quasar CTS 30.10. The colour-coded regions represent the result of $\chi^2$ fitting with the star denoting the minimum and the orange and the brown areas stand for 1$\sigma$ and 2$\sigma$ uncertainties, respectively. \textbf{Left panel:} Time-delay of $562 \pm 22$ days. \textbf{Right panel:} Time-delay of $564 \pm 2$ days as determined by the JAVELIN code.}
    \label{fig_CTS_cosmology}
\end{figure}

In particular, \citet{2019ApJ...880...46C} and \citet{2019arXiv190703910Z} applied six different methods to determine the time-lag between the continuum and MgII line response. These included cross-correlation based techniques ICCF, DCF, zDCF, $\chi^2$-based method \citep{2013A&A...556A..97C,2019ApJ...880...46C}, the measure of data regularity \citep[von Neumann's estimator;][]{2017ApJ...844..146C}, and the JAVELIN code \citep[Just Another Vehicle for Estimating Lags In Nuclei, formerly SPEAR;][]{2011ApJ...735...80Z,2013ApJ...765..106Z,2016ApJ...819..122Z}. They showed that for the intermediate-redshift quasar CTS C30.10 (see also Section~\ref{subsec_time_lag}), the uncertainty in the time-lag determination depends on the approach. If a narrow surrounding of the main peak in time delay measure is taken into consideration, the uncertainty can be of the order of a few days only for the total observing run of 1000 days and more. The small uncertainty of the order of a few days only was obtained by JAVELIN. On the other hand, if the total range of possible time delays is considered, that is between zero and half of the duration of the observation, the uncertainty can be of the order of 10 and even 100 days depending on the quality of the data. This uncertainty further propagates into the radius-luminosity relation and the Hubble diagram of reverberation mapped quasars. In Fig.~\ref{fig_CTS_cosmology}, we illustrate this in terms of constraining cosmological parameters from monitoring of a single source CTS C30.10 mentioned above (dark matter and dark energy fractions) for two different uncertainties -- 2 days as provided by the JAVELIN code and 22 days that is more representative when other methods are taken into account.

Although the error provided by the JAVELIN is underestimated and can be corrected by performing many bootstrap runs to recover the overall time-lag distribution, by combining several robust methods one can achieve the time-lag uncertainty of $5\%$. Under this assumption, approximately 625 sources are needed to achieve $2\% $ uncertainty for all cosmological parameters including $w_0$ parameter. However, special attention is needed for the process of a systematic time-lag analysis for the whole sample of reverberation-mapped quasars since currently, the time-lag analysis is largely heterogeneous with different authors using different time-delay methods and their associated uncertainties. This is a source of large systematic uncertainties that will need to be addressed in future RM samples of quasars.  

Quasars are complex objects and many corrections are still needed to consider these objects like standard candles suitable for cosmology. However, Cepheids star and SNIa have been also corrected for the shape of the lightcurve, extinction and the host galaxy, and sophisticated observational methods have been proposed for getting better observations \cite[e.g.][]{scolnic2018}. All this effort is reflected in the low uncertainties associated with the cosmological results. A similar situation should be happening with quasars taking advantage of their broad range of redshift and luminosity.

\section{Photometric Reverberation Mapping in the LSST era}
\label{sec_LSST}
\subsection{AGN Science with LSST}

\textit{LSST} will be a public optical/NIR survey of $\sim$half the sky in the \textit{ugrizy} bands upto r$\sim$27.5 based on $\sim820$ visits across all the six photometric bands over a 10-year period \citep{lsst2019}. The 8.4m telescope with a \textit{state-of-the-art} 3.2 Gigapixel flat-focal array camera will allow performing rapid scans of the sky with 15 seconds exposure and thus providing a moving array of color images of objects that change. The whole observable sky is planned to be scanned every $\sim$4 nights. It is expected that upon the completion of the main-survey period, \textit{LSST} will have mapped $\sim$20 billion galaxies and $\sim$ 17 billion stars using these six photometric bands\footnote{see \citet{lsst2019} for a complete review on \textit{LSST} science drivers, telescope design and data products}. 

The \textit{LSST} AGN survey will produce a high-purity sample of at least 10 million well-defined, optically-selected AGNs (see Section \ref{sub:suitability}). Utilizing the large sky coverage, depth, the six filters extending to 1$\rm{\mu}$m, and the valuable temporal information of \textit{LSST}, this AGN survey will supersede the largest current AGN samples by more than an order of magnitude. Each region of the \textit{LSST} sky will receive $\sim$200 visits in each band in the decade long monitoring, allowing variability to be explored on timescales from minutes to a decade (see Chapter 10 of \citet{lsst_book} for more details). \textit{LSST} will conduct more intense observations of at least 4 \textit{Deep Drilling Fields} or DDFs, i.e. COSMOS, XMM-LSS, W-CDF-S and ELAIS-S1, each of which cover $\sim$10 deg$^2$ (more details in \citealt{lsst_wp}). \citet{lsst_wp} estimates an increase in the depth in the DDFs by over an order of magnitude relative to the main-survey. This will be made possible owing to a $\sim$18-20X increase in the number of visits per photometric band.

\textit{LSST} will be assisted by an array of campaigns \citep{brandt2017} during its proposed run. Moreover, there is a need for follow-up spectroscopic campaigns i.e. SDSS-V Black Hole Mapper \citep{SDSS5}, which aim to derive BLR properties and reliable SMBH masses for distant AGNs with expected observed-frame reverberation lags of 10–1000 days. The SDSS-V Black Hole Mapper survey will also perform reverberation mapping campaigns in three out of the four \textit{LSST} DDFs. \textit{LSST} will provide additional high quality photometry for these sources that will substantially improve these estimations. The proposed cadence ($\sim$2 nights) of the observations of AGNs and the expected re-visits during the 10-year run will allow to (1) ensure a high lag recovery fraction for the relatively short accretion-disk lags expected \citep{Yu2018}; and (2) allow investigations of secular evolution of these lags that test the underlying model. \textit{LSST} will also perform quality accretion disk reverberation mapping for $\sim$3000 AGNs in the DDFs.

\subsection{Suitability of quasar monitoring}
\label{sub:suitability}
Every night, \textit{LSST} will monitor $\sim$75 million AGNs and is estimated to detect $\sim$300+ million AGNs in the $\sim$18000 deg$^2$ main-survey area \citep{Luo2017}. Taking into account factors like obscuration and host-galaxy contamination that will hinder this AGN selection, optimistic estimates predict around $\sim$20 million (from \textit{LSST} alone) and $\sim$50+ million (\textit{LSST}+ others) AGNs to be reliably monitored.

\begin{figure}[h!]
\begin{center}
\includegraphics[width=\textwidth]{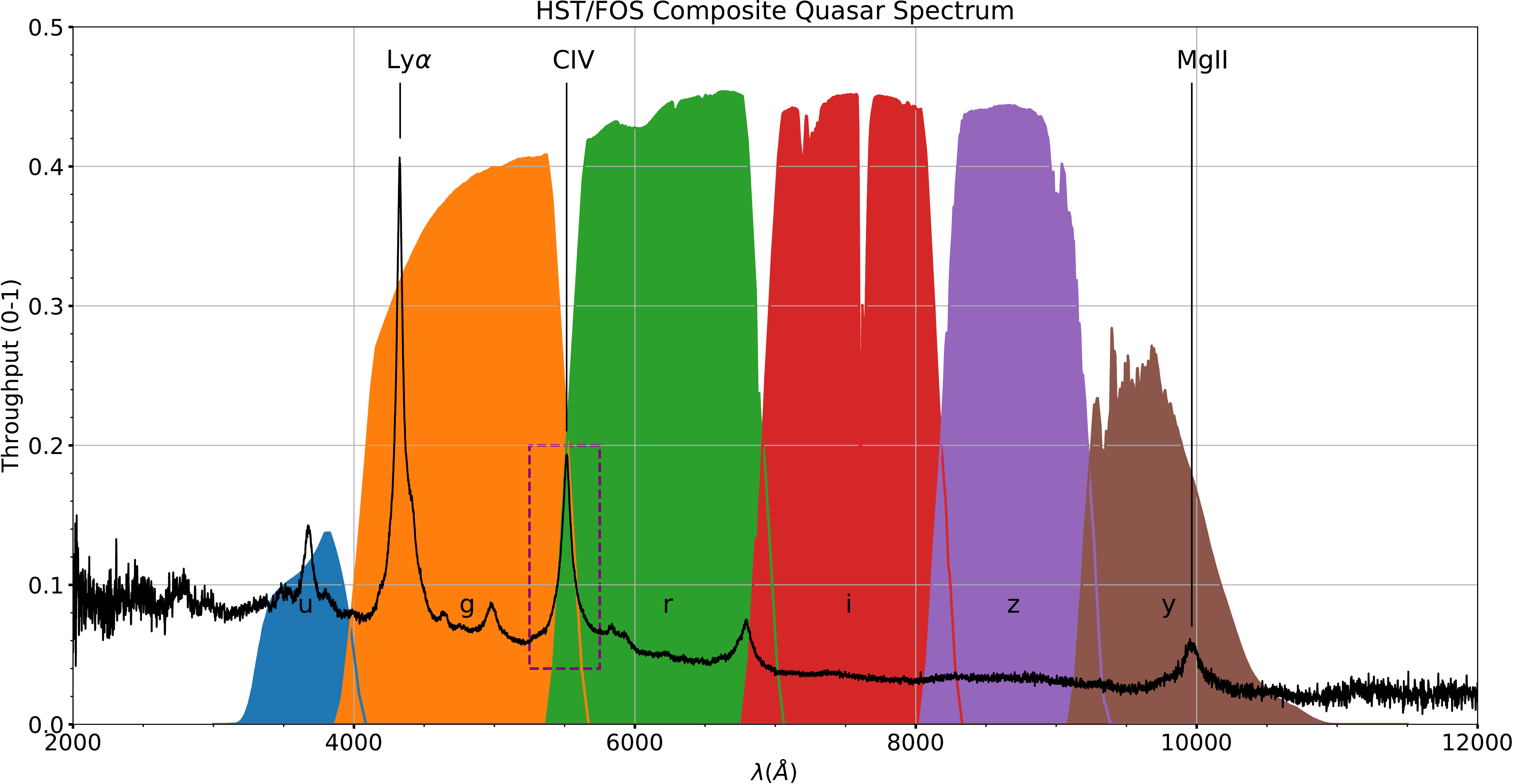}
\end{center}
\caption{Throughput curves for the 6 \textit{LSST} photometric bands (\textit{ugrizy}), shown as a function of wavelength (in \AA). The black solid line is the \textit{HST/FOS} composite quasar spectrum \citep{zheng1997} and the dashed box in purple highlights the importance of \textit{quasar selection}. The spectrum is shown at a redshift, \textit{z} = 2.56. The composite spectrum is downloaded from \href{https://archive.stsci.edu/prepds/composite_quasar/}{https://archive.stsci.edu/prepds/composite\_quasar/}}\label{fig:lsst_bands}
\end{figure}

\subsubsection{Type-1 quasar counts}
\label{quasar-counts}
In the context of the reverberation mapping, one intends to focus on the broad emission lines that are a characteristics of Type-1 AGNs \citep{1997ASSL..218...85N, 2004ApJ...613..682P, 2018Natur.563..657G}. With \textit{LSST}, we would want to constrain the quasar counts in terms of only Type 1 AGNs with respect to the observed broad emission lines such as H$\beta$, MgII, CIV. We perform a simple filtering to estimate these numbers in terms of total predicted number of quasars to be observed over the full duration of \textit{LSST}. 

\begin{figure}[h!]
\begin{center}
\includegraphics[width=0.85\textwidth]{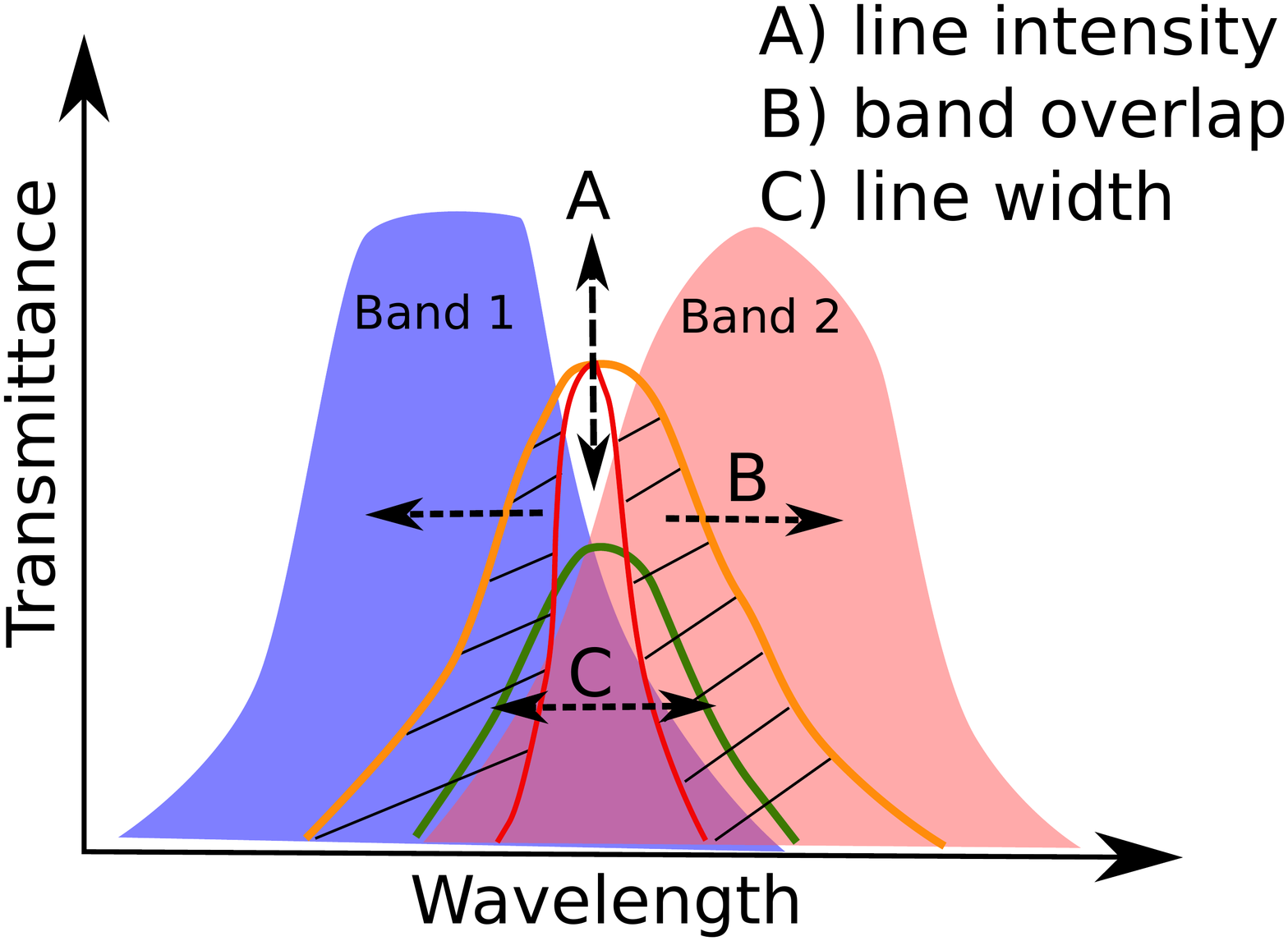}
\end{center}
\caption{A representative illustration for the quasar selection based on distribution of (A) line intensities; (B) band overlaps, and (c) line widths. Three instances of emission line profiles are shown in orange, red and green, and the patched region highlights the importance of the broad distribution of equivalent widths (EWs) in quasars. The arrows mark the direction of the effect due to these three factors. }\label{fig:lsst_illustration}
\end{figure}

Three entities affect our estimation of the ``good'' Type 1 quasars, namely (1) the line intensity; (2) overlap of the photometric bands; and (3) the line FWHM. The peak intensity of the line (including continuum) that will be acquired from the pipeline, is already convolved with the filters and to recover the true intensity one needs to scale by $\frac{1}{T}$, where \textit{T} is the throughput (or transmittance) of the corresponding band which is estimated from the band function. We adopt a throughput value of 0.2 to exclude the gaps between the consecutive bands (see Figure \ref{fig:lsst_bands}). To get an idea of the impact of the line FWHM on the quasar number counts, we adopted the mean values for the broad emission lines -- H$\beta$, MgII, from the SDSS DR7 Quasar catalogue \citep{shen11}. The average values for the FWHM are: $4662.26\pm2181.53\;\rm{km\;s}^{-1}$ (MgII) and $4903.63\pm3119.26\;\rm{km\;s}^{-1}$ (H$\beta$). For simplicity, we adopt the values 4500 km s$^{-1}$ (MgII) and 5000 km s$^{-1}$ (H$\beta$) for this preliminary analysis. Since, the emission lines in consideration are ``broad'', we estimate the equivalent width of MgII and H$\beta$ to account for the broadness in these line profiles. We estimate this additional width from the mean FWHMs and the values obtained are $\sim$42 \AA\ and $\sim$80 \AA\ for MgII and H$\beta$ respectively. Next, we retrieve the wavelength windows where these photometric bands overlap (see Table \ref{tab: table1}). As we have assumed the throughput value of 0.2, we exclude the u-band since the peak transmittance in this band is $\sim$0.14. This is the reason why the first entry in Table \ref{tab: table1} is in the g-band (for MgII; rest-frame wavelength = 2799.12\AA) and the wavelength is recovered corresponds to the intersection with the throughput value, \textit{T}=0.2 on the lower wavelength side of the band. In the case for H$\beta$, the starting value z=0 is inclusive (rest-frame wavelength for H$\beta$ = 4861.36\AA). The remaining entries in the Table \ref{tab: table1} are identical. We estimate that $\sim$26\% (for MgII) and $\sim$36\% (for H$\beta$) of the supposed Type 1 quasars will end up being observed in these overlapping windows taking into account the respective emission lines broadening and our naive assumption of the throughput cutoff. Figure \ref{fig:lsst_illustration} shows an illustrative view of this analysis in a more general context.

The expected number of Type 2 quasars is estimated to be $\sim$50-70\% of the total population \citep{schmitt2001, hao2005, elitzur2012}. So, if we start with a total population, for example, 1 million, we will have about 300,000-500,000 Type 1 AGNs based on this filtering. Next, the actual fraction of the ``good'' quasars will be $\sim$22-37\% (MgII) and $\sim$19-32\% (H$\beta$) after applying our filter due to the band overlap. It is expected that the filter wheel to be installed on \textit{LSST} will have five slots (the filter wheel will prioritize \textit{griz} and the \textit{y} and \textit{u} filters will be alternated for the fifth slot) complementary to the SDSS filter design \citep{1996AJ....111.1748F}. This means that at any given time, the image will be taken with one of the filters in-line with the camera. The filter replacement takes about 2 minutes to complete and this filter replacement will be performed 2-6 times per night and the total number of filter changes through the survey is 14,194 \citep{2017arXiv170804058L}.  The remaining filter will be substituted for one of the existing filters in the filter-wheel (most likely u-band in place of y-band). This will affect the quasar counts based on the quality of the data, where quality is directly related to the number of re-visits for the source.

\begin{table}[]
\centering
\caption{LSST photometric bands: overlapping windows}
\vspace{0.25cm}
\label{tab: table1}
\begin{tabular}{ccc}
\multicolumn{1}{c}{} & $\lambda_{\rm{min}}$ (\AA) & $\lambda_{\rm{max}}$ (\AA)\\ \hline\hline
\multicolumn{1}{c|}{g$_{\rm{min}}$} & -- & 4091$^{a}$, 4861.33$^{b}$\\ \hline
\multicolumn{1}{c|}{g,r} & 5370 & 5669\\ \hline
\multicolumn{1}{c|}{r,i} & 6760 & 7059\\ \hline
\multicolumn{1}{c|}{i,z} & 8030 & 8329\\ \hline
\multicolumn{1}{c|}{z,y} & 9084 & 9385\\ \hline
\multicolumn{1}{c|}{y$_{\rm{max}}$} & 9894 & -- \\ \hline
\footnotesize{$^a$ MgII, $^b$ H$\beta$}
\end{tabular}
\end{table}

\subsection{Modelling `real' light curves}
\textit{LSST} is a photometric project but the 6-channel photometry (see Figure \ref{fig:lsst_bands}) can be effectively used for the purpose of reverberation mapping and estimation of time delays. We present some preliminary results from our software in development which allows to produce \textit{mock} light curves and recover the time delays. The code takes into consideration several key parameters to produce these light curves, namely -- (1) the campaign duration of the instrument (10 years); (2) number of visits per photometric band; (3) the photometric accuracy (0.01-0.1 mag)\footnote{these values are adopted from \citet{lsst2019}}; (4) black hole mass distribution\footnote{the results shown here are for a representative black hole mass, M$_{\rm{BH}}$ = 10$^8$ M$_{\odot}$.}; (5) luminosity distribution\footnote{the results shown here are for two representative cases of bolometric luminosity, L$_{\rm{bol}}$ = 10$^{45}$ and 10$^{46}$ erg s$^{-1}$.}; (6) redshift distribution\footnote{the results shown here are for two representative cases of redshifts, z = 0.1 and 0.985.}; and (7) line equivalent widths (EWs) consistent with SDSS quasar catalogue \citep{shen11}. We create continuum stochastic lightcurve for a quasar of an assumed magnitude and redshift from AGN power spectrum with Timmer-Koenig algorithm \citep{tk1995}. The code takes as an input a first estimate for the time delay measurement. We utilize the standard \RL\ relation \citep{bentz2013} to estimate this value. In the current version of the code, the results for the photometric reverberation method are estimated by adopting only 2 photometric channels at a time and the time delay is estimated using the $\chi^2$ method. We account for the contamination in the emission line (H$\beta$, MgII, CIV) as well as the in the continuum. The code also incorporates the FeII pseudo-continuum and contamination from starlight i.e. stellar contribution.

\begin{figure}[h!]
\begin{center}
\includegraphics[width=\textwidth]{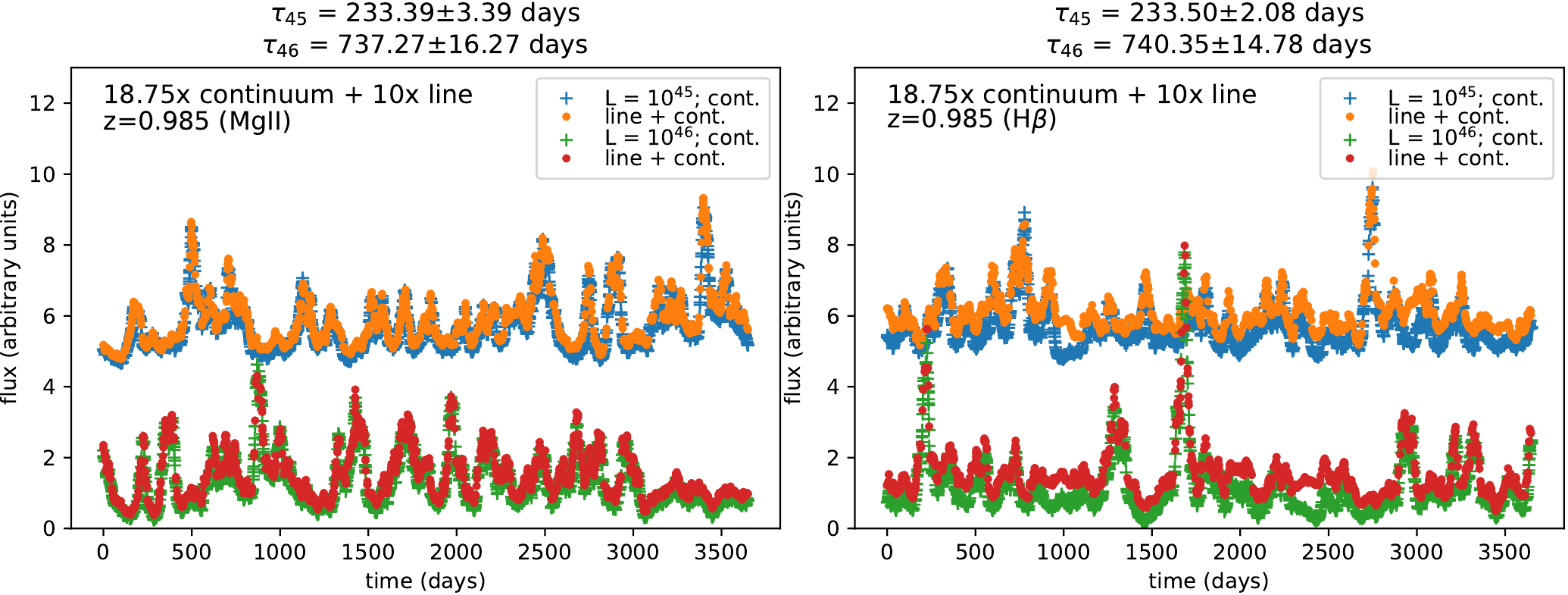}
\end{center}
\caption{Simulated lightcurves for \MgII\ and H$\beta$ at \textit{z} = 0.985. The lightcurves are generated using \textit{Timmer-Koenig} method \citep{tk1995} and combining with photometric data from \textit{LSST} \citep{lsst2019}. Two cases of luminosity ($10^{45}$ and $10^{46}$ erg s$^{-1}$) are shown, each for the continuum and the line. The continuum and the line contributions are made denser by a factor of 18.75 and 10, respectively. The lightcurves with luminosity $10^{45}$ erg s$^{-1}$ is shifted by factor of 4.5 with respect to the lightcurves with luminosity $10^{46}$ erg s$^{-1}$ to highlight the differences between the two cases. The corresponding time delays are reported on top of each plot for these two cases of luminosity. The simulations are currently performed using two channels.}\label{fig:lsst_lc} 
\end{figure}

In \citet{2019arXiv191002725L}, we show preliminary results from our code. In that paper, we show the variation in the simulated lightcurves for MgII and H$\beta$ as a function of the redshift and luminosity and their corresponding time-delay distributions. The power spectral distribution for these lightcurves was assumed to have the low-frequency break corresponding to 2000 days. In Figure \ref{fig:lsst_lc}, we show the results from subsequent analysis performed using two-channel setup. We incorporate a higher density of the continuum and the the line contribution owing to the fact that in the DDFs the probing will be denser compared to the rest of the All-Sky survey. The computations are performed for a representative case of black hole mass, $M_{BH} = 3\times 10^8\;M_{\odot}$. The photometric accuracy is kept at 0.01 and the low-frequency break corresponds to between 200-300 days. The EWs are increased by a factor of 3 with respect to the average values (MgII: 47 \AA\ and H$\beta$: 87 \AA) which were used in the previous results. The corresponding time-delays (with dispersion) recovered from the analyses are reported for the respective emission line (MgII and H$\beta$) and two representative cases of the luminosity, $L_{3000} = 10^{45}\;$erg s$^{-1}$ and $10^{46}\;$erg s$^{-1}$. The dispersions for the reported time-delays are found to be within $\sim$1-2\% limit. 

\section{Discussions}
\label{section_discussion}

The MgII and H$\beta$ both are low-Ionization broad-emission lines, however, MgII shows a narrower profile than \hb\ profile \citep{2013A&A...555A..89M}, suggesting that the emission line is emitted in an outer region. In \cite{snieg2019}, we have tested that their emissivity profiles using CLOUDY reveals similar emitting zones in the BLR. When compared to their corresponding FeII emission, the maximum MgII and H$\beta$ emission peaks are $\sim$ 10$^{7}$ cm deeper from the face of the BLR cloud. Yet, these emitting zones are extended, spanning across $\sim$6 orders in cloud depth. There have been recent studies that predict that the most efficient MgII emitting clouds are always near the outer physical boundary of the BLR, while the Balmer line gas is inside this outer BLR  boundary \citep{guo2019}. This explains the low variability shown by MgII in comparison to \hb. Since MgII usually does not display a ``breathing mode'', it suggests that there is not a Radius-Luminosity relation for this line. However, the possibility of a global Radius-Luminosity is viable, as long as the outer radius of the BLR scales with the black hole mass \citep{guo2019}. The Radius-Luminosity relation for MgII (Figure \ref{fig_RL_MgII}) shows a slope of $\alpha=0.59\pm 0.06$, which is in agreement with the idea of a general R-L relation.

As has been shown in this paper, the Eddington ratio plays an important role in the \RL{}. This is straightforward as we see that if the line width is fixed but we recover a lower time delay than expected and subsequently the black hole mass estimated is lower. Hence, the Eddington ratio that we estimate will be higher. This trend of decrease of the time delay with rising Eddington ratio seems systematic \citep{2017ApJ...851...21G, martinezaldama2019}. Being able to constrain their physical parameter space (in terms of the ionisation parameter, cloud density and accretion rate) and link this to the observables, such as emission line ratios and shape of the emission line profiles, would provide answers to this anomaly (Panda et al. in prep).

The next steps in the code development for modeling the light curves, include testing with different methodologies to generate lightcurves e.g. Damped-Random Walk. This also applies to the time-delay estimation where we are testing the consistency of the predicted values with other available methods e.g. Interpolated cross-correlation function \citep{1987ApJS...65....1G} and JAVELIN \citep{2011ApJ...735...80Z,2013ApJ...765..106Z,2016ApJ...819..122Z}. In the near future, we plan to extend the analysis taking into account contribution from all the 6 channels, including the lag between the images when going from one filter to the next and including the filter replacement time as explained in Sec. \ref{quasar-counts}. Finally, we plan to create lightcurves incorporating a distribution of the redshift, luminosity, $M_{BH}$ and EWs.

A special attention should be paid to realistic estimates of the uncertainties of different time-lag methods \citep[ICCF, DCF, zDCF, JAVELIN, $\chi^2$, regularity measures-von Neumann estimator, see][for a general overview and the comparison]{2019arXiv190703910Z} since this is one of the main sources of the scatter of the radius-luminosity relation. The effect of systematic errors on the time-lag uncertainty was analyzed by \citet{2019arXiv190903072Y}, who compared JAVELIN and ICCF methods using simulated light curves, with JAVELIN performing better in terms of uncertainty determination. A similar result is reached by \citet{2019arXiv190903092I} who compare ICCF, JAVELIN, and zDCF methods using generated light curves that mock multi-object spectroscopic reverberation mapping (MOS-RM) surveys. They conclude that JAVELIN and ICCF outperform zDCF, with JAVELIN generally introducing the lowest bias into the radius-luminosity relation. However, more comparison and analysis is needed using both the observed reverberation-mapped quasars (e.g. bootstrap algorithm) as well as simulated light curves (generated by both Timmer-Koenig algorithm and damped random walk for comparison) for the traditional (ICCF, DCF, zDCF) and other, novel methods of the time-lag determination (von Neumann, $\chi^2$ methods), with the special focus on how time-delay uncertainties propagate into constraining cosmological parameters \citep{2013A&A...556A..97C}.

\section{Conclusions}
\label{section_conclusions}

Reverberation mapping studies have been quite successful to constrain the $R_{\rm{BLR}} - L_{\lambda}$ relation for various broad emission lines (H$\beta$: \citealt{2017ApJ...851...21G} and references therein; MgII: \citealt{2006ApJ...647..901M, 2013A&A...556A..97C, 2018ApJ...865...56L, 2019ApJ...880...46C}; CIV: \citealt{2019arXiv190403199G} and references therein). In this contribution, we present the monitoring of the intermediate-redshift quasar CTS C30.10 by the SALT telescope over the course of six years and detected the time-delay of MgII-line response of $\tau_{MgII}=551^{+49}_{-82}$ days in the source frame. In combination with other sources where MgII time-delay was detected, we constructed the radius-luminosity relation considering a sample of 117 sources taken from the literature, and fitted it with the general power-law relation $\log(R_{\rm BLR}/1 \text{lt-day})=K+\alpha\log{(\lambda L_{\lambda}/10^{44}\,{\rm erg\,s^{-1}})}$, with the best-fit coefficients of $K=1.47 \pm 0.06$ and $\alpha=0.59\pm 0.06$, which are within uncertainties consistent with the values of \RL\, relation by \citet{bentz2013}. It is thus possible to use the radius-luminosity relation also towards higher redshifts using MgII and potentially CIV line. Under the assumption that the time-delay can be measured within $5\%$ uncertainty, about 625 sources are needed in future surveys to constrain the cosmological parameters with 2\% uncertainty. 

Along \RL\ relation there is an effect of the accretion rate, which induces a departure from the expected value, i.,e., super-Eddington sources show a time delay shorter than the expected. This effect can be corrected, recovering the classical low scatter in the relation. With the estimation of the luminosity distance, we estimated the cosmological parameters. The results are in agreement with the $\Lambda$CDM model within 2$\sigma$ confidence level, which is still not suitable for cosmological results. Unfortunately, there are still some systematic errors that have to be corrected. The current sample size for such studies has now reached $\gtrsim$100 and future reverberation campaigns promise to increase this sample by manifolds and get accurate cosmological results. Also,  an improvement in the method for the determination of the time delay and a better understanding of the AGN properties (e.g. inclination angle) are required to decrease the errors and get better results \citep{panda19b}.

\textit{LSST} is one of the forerunners in such future campaigns which will cover a wide range of wavelength (3050 \AA $\lesssim \lambda \lesssim$ 11000 \AA) and this instrument alone will provide $\sim$5 orders increase in the number of reverberation-mapped quasars. With \textit{LSST} we will have no dearth in the number of quasars, yet if we are to find answers to some of the long-standing questions in astronomy with the help of quasars, we would require to improve the quality of the observed data from the instrument. In this paper, we provide the foundation for a filtering algorithm that will be useful for the community to account for the ``good'' quasars especially for the purpose of variability studies and photometric reverberation mapping. The suitability of these good quasars is based on three fundamental criteria -- the emission line intensity, the band overlap and the line width. We also show the first results from our code designed to create synthetic, stochastic lightcurves incorporating fundamental quasar properties and the instrument's performance as per \cite{lsst2019}. This will be useful -- first for preparing a mock catalog of quasar lightcurves, and second, to test real data that will be obtained from the \textit{LSST} in the near future.

\section*{Conflict of Interest Statement}

The authors declare that the research was conducted in the absence of any commercial or financial relationships that could be construed as a potential conflict of interest.


\section*{Funding}
The project was partially supported by National Science Centre, Poland, grant No. 2017/26/A/ST9/00756 (Maestro 9), and by the MNiSW grant DIR/WK/2018/12.

\section*{Acknowledgments}

We cordially thank all the organizers of Symposium 2 ``Quasars in cosmology'' organized at EWASS 2019 in Lyon on June 25-26, 2019, in particular Bo\.zena Czerny, Paola Marziani, Edi Bon, Natasha Bon, Elisabeta Lusso, and Mauro D'Onofrio.



\bibliographystyle{frontiersinSCNS_ENG_HUMS} 
\bibliography{article}





\end{document}